\documentclass[11pt,a4paper,]{article}
\usepackage{lmodern}

\usepackage{amssymb,amsmath}
\usepackage{ifxetex,ifluatex}
\usepackage{fixltx2e} % provides \textsubscript
\ifnum 0\ifxetex 1\fi\ifluatex 1\fi=0 % if pdftex
  \usepackage[T1]{fontenc}
  \usepackage[utf8]{inputenc}
\else % if luatex or xelatex
  \usepackage{unicode-math}
  \defaultfontfeatures{Ligatures=TeX,Scale=MatchLowercase}
\fi
\IfFileExists{upquote.sty}{\usepackage{upquote}}{}
\IfFileExists{microtype.sty}{%
\usepackage[]{microtype}
\UseMicrotypeSet[protrusion]{basicmath} % disable protrusion for tt fonts
}{}
\PassOptionsToPackage{hyphens}{url} % url is loaded by hyperref
\usepackage[unicode=true]{hyperref}
\hypersetup{
            pdftitle={Conditional normalization in time series analysis},
            pdfkeywords={conditional normalization, missing value imputation, conditional autocorrelation, conditional cross-correlation, lag time estimation, stream data, water quality},
            pdfborder={0 0 0},
            breaklinks=true}
\usepackage{geometry}
\usepackage[style=authoryear-comp,]{biblatex}
\usepackage{longtable,booktabs}
\IfFileExists{footnote.sty}{\usepackage{footnote}\makesavenoteenv{long table}}{}
\IfFileExists{parskip.sty}{%
\usepackage{parskip}
}{% else
\setlength{\parindent}{0pt}
\setlength{\parskip}{6pt plus 2pt minus 1pt}
}
\providecommand{\tightlist}{%
  \setlength{\itemsep}{0pt}\setlength{\parskip}{0pt}}
\makeatletter
\def\fps@figure{htbp}
\makeatother

\title{Conditional normalization in time series analysis}

\RequirePackage{caption}
\DeclareCaptionStyle{italic}[justification=centering]
 {labelfont={bf},textfont={it},labelsep=colon}
\RequirePackage{bera}
\RequirePackage{mathpazo}

\RequirePackage{fancyhdr}
\fancypagestyle{plain}{%
\fancyhf{} % clear all header and footer fields
\fancyfoot[C]{\sffamily\thepage} % except the center

}

\RequirePackage{bm,amsmath}
\allowdisplaybreaks

\RequirePackage{graphicx}
\RequirePackage[compact,sf,bf]{titlesec}
\titleformat{\section}[block]
  {\fontsize{15}{17}\bfseries\sffamily}
  {\thesection}
  {0.4em}{}
\titleformat{\subsection}[block]
  {\fontsize{12}{14}\bfseries\sffamily}
  {\thesubsection}
  {0.4em}{}
\titlespacing{\section}{0pt}{*5}{*1}
\titlespacing{\subsection}{0pt}{*2}{*0.2}

\def\Date{\number\day}
\def\Month{\ifcase\month\or
 January\or February\or March\or April\or May\or June\or
 July\or August\or September\or October\or November\or December\fi}
\def\Year{\number\year}

\makeatletter
\def\wp#1{\gdef\@wp{#1}}\def\@wp{??/??}
\def\jel#1{\gdef\@jel{#1}}\def\@jel{??}
\def\showjel{{\large\textsf{\textbf{JEL classification:}}~\@jel}}

\def\addresses#1{\gdef\@addresses{#1}}\def\@addresses{??}
\def\cover{{\sffamily\setcounter{page}{0}
        \thispagestyle{empty}
        \placefig{2}{1.5}{width=5cm}{monash2}
        \placefig{16.9}{1.5}{width=2.1cm}{MBusSchool}
        \begin{textblock}{4}(16.9,4)ISSN 1440-771X\end{textblock}
        \begin{textblock}{7}(12.7,27.9)\hfill
        \includegraphics[height=0.7cm]{AACSB}~~~
        \includegraphics[height=0.7cm]{EQUIS}~~~
        \includegraphics[height=0.7cm]{AMBA}
        \end{textblock}
        \vspace*{2cm}
        \begin{center}\Large
        Department of Econometrics and Business Statistics\\[.5cm]
        \footnotesize http://monash.edu/business/ebs/research/publications
        \end{center}\vspace{2cm}
        \begin{center}
        \fbox{\parbox{14cm}{\begin{onehalfspace}\centering\Huge\vspace*{0.3cm}
                \textsf{\textbf{\expandafter{\@title}}}\vspace{1cm}\par
                \LARGE\@author\end{onehalfspace}
        }}
        \end{center}
        \vfill
                \begin{center}\Large
                \Month~\Year\\[1cm]
                Working Paper \@wp
        \end{center}\vspace*{2cm}}}
\def\pageone{{\sffamily\setstretch{1}%
        \thispagestyle{empty}%
        \vbox to \textheight{%
        \raggedright\baselineskip=1.2cm
     {\fontsize{24.88}{30}\sffamily\textbf{\expandafter{\@title}}}
        \vspace{2cm}\par
        \hspace{1cm}\parbox{14cm}{\sffamily\large\@addresses}\vspace{1cm}\vfill
        \hspace{1cm}{\large\Date~\Month~\Year}\\[1cm]
        \hspace{1cm}\showjel\vss}}}
\def\blindtitle{{\sffamily
     \thispagestyle{plain}\raggedright\baselineskip=1.2cm
     {\fontsize{24.88}{30}\sffamily\textbf{\expandafter{\@title}}}\vspace{1cm}\par
        }}
\def\titlepage{{\cover\newpage\pageone\newpage\blindtitle}}

\def\blind{\def\titlepage{{\blindtitle}}\let\maketitle\blindtitle}
\def\titlepageonly{\def\titlepage{{\pageone\end{document}}}}
\def\nocover{\def\titlepage{{\pageone\newpage\blindtitle}}\let\maketitle\titlepage}
\let\maketitle\titlepage
\makeatother

\RequirePackage{setspace}
\spacing{1.5}

\RequirePackage{microtype}

\RequirePackage{xcolor} % Needed for links
\definecolor{darkblue}{rgb}{0,0,.6}
\RequirePackage{url}

\makeatletter
\@ifpackageloaded{hyperref}{}{\RequirePackage{hyperref}}
\makeatother
\hypersetup{
     citecolor=0 0 0,
     breaklinks=true,
     bookmarksopen=true,
     bookmarksnumbered=true,
     linkcolor=darkblue,
     urlcolor=blue,
     citecolor=darkblue,
     colorlinks=true}

\newenvironment{keywords}{\par\vspace{0.5cm}\noindent{\sffamily\textbf{Keywords:}}}{\vspace{0.25cm}\par\hrule\vspace{0.5cm}\par}

\renewenvironment{abstract}{\begin{minipage}{\textwidth}\parskip=1.4ex\noindent
\hrule\vspace{0.1cm}\par{\sffamily\textbf{\abstractname}}\newline}
  {\end{minipage}}

\usepackage[T1]{fontenc}
\usepackage[utf8]{inputenc}

\usepackage[showonlyrefs]{mathtools}
\usepackage[no-weekday]{eukdate}

\makeatletter
\@ifpackageloaded{biblatex}{}{\usepackage[style=authoryear-comp, backend=biber, natbib=true]{biblatex}}
\makeatother
\DeclareFieldFormat{url}{\texttt{\url{#1}}}
\DeclareFieldFormat[article]{pages}{#1}
\DeclareFieldFormat[inproceedings]{pages}{\lowercase{pp.}#1}
\DeclareFieldFormat[incollection]{pages}{\lowercase{pp.}#1}
\DeclareFieldFormat[article]{volume}{\mkbibbold{#1}}
\DeclareFieldFormat[article]{number}{\mkbibparens{#1}}
\DeclareFieldFormat[article]{title}{\MakeCapital{#1}}
\DeclareFieldFormat[inproceedings]{title}{#1}
\DeclareFieldFormat{shorthandwidth}{#1}
\usepackage{xpatch}
\xpatchbibmacro{volume+number+eid}{\setunit*{\adddot}}{}{}{}
\renewbibmacro{in:}{%
  \ifentrytype{article}{}{%
  \printtext{\bibstring{in}\intitlepunct}}}

\makeatletter
\DeclareDelimFormat[cbx@textcite]{nameyeardelim}{\addspace}
\makeatother

\wp{no/yr}
\jel{C10,C14,C22}

\RequirePackage[absolute,overlay]{textpos}
\def\placefig#1#2#3#4{\begin{textblock}{.1}(#1,#2)\rlap{\includegraphics[#3]{#4}}\end{textblock}}

\nocover

\author{Puwasala~Gamakumara, Edgar~Santos-Fernandez, Priyanga Dilini~Talagala, Rob J~Hyndman, Kerrie~Mengersen, Catherine~Leigh}
\addresses{\textbf{Puwasala Gamakumara}\newline
Department of Econometrics \& Business Statistics\newline Monash University, Clayton VIC 3800, Australia
\newline{Email: \href{mailto:puwasala.gamakumara@gmail.com}{\nolinkurl{puwasala.gamakumara@gmail.com}}}\newline Corresponding author\\[1cm]
\textbf{Edgar Santos-Fernandez}\newline
School of Mathematical Sciences\newline Queensland University of Technology, Brisbane QLD 4000, Australia
\newline{Email: \href{mailto:edgar.santosfernandez@qut.edu.au}{\nolinkurl{edgar.santosfernandez@qut.edu.au}}}\\[1cm]
\textbf{Priyanga Dilini Talagala}\newline
Department of Computational Mathematics, University of Moratuwa, Bandaranayake Mawatha, Moratuwa, 10400, Sri Lanka
\newline{Email: \href{mailto:priyangad@uom.lk}{\nolinkurl{priyangad@uom.lk}}}\\[1cm]
\textbf{Rob J Hyndman}\newline
Department of Econometrics \& Business Statistics\newline Monash University, Clayton VIC 3800, Australia
\newline{Email: \href{mailto:Rob.Hyndman@monash.edu}{\nolinkurl{Rob.Hyndman@monash.edu}}}\\[1cm]
\textbf{Kerrie Mengersen}\newline
Science and Engineering Faculty \newline School of Mathematical Sciences \newline Queensland University of Technology, Brisbane QLD 4000, Australia
\newline{Email: \href{mailto:k.mengersen@qut.edu.au}{\nolinkurl{k.mengersen@qut.edu.au}}}\\[1cm]
\textbf{Catherine Leigh}\newline
Biosciences and Food Technology Discipline \newline School of Science\newline RMIT University, Bundoora VIC 3082, Australia
\newline{Email: \href{mailto:catherine.leigh@rmit.edu.au}{\nolinkurl{catherine.leigh@rmit.edu.au}}}\\[1cm]
}

\date{\sf\Date~\Month~\Year}

 \usepackage{todonotes}    

\begin{document}
\maketitle
\begin{abstract}
Time series often reflect variation associated with other related variables. Controlling for the effect of these variables is useful when modeling or analysing the time series. We introduce a novel approach to normalize time series data conditional on a set of covariates. We do this by modeling the conditional mean and the conditional variance of the time series with generalized additive models using a set of covariates. The conditional mean and variance are then used to normalize the time series. We illustrate the use of conditionally normalized series using two applications involving river network data. First, we show how these normalized time series can be used to impute missing values in the data. Second, we show how the normalized series can be used to estimate the conditional autocorrelation function and conditional cross-correlation functions via additive models. Finally we use the conditional cross-correlations to estimate the time it takes water to flow between two locations in a river network.
\end{abstract}
\begin{keywords}
conditional normalization, missing value imputation, conditional autocorrelation, conditional cross-correlation, lag time estimation, stream data, water quality
\end{keywords}

\hypertarget{introduction}{%
\section{Introduction}\label{introduction}}

Normalization of some variables is often required prior to using a statistical or machine learning algorithm. In this study, we introduce a novel normalization method for time series data which aims primarily to remove the conditional variation in the time series that is induced by other sources of variation.

Common data normalization methods, such as min-max transformation or standardization (also called z-score normalization) are not always applicable for time series data. The normalizing constants may change in the future, and they assume stationary processes. For non-stationary data, sliding-window normalization has been proposed \autocites[e.g.,][]{ogasawara2010,vafaeipour2014}, where the time series is divided into windows of a specified length and data are normalized within each window. However, these methods do not account for any external variables that can influence the variation in the time series.

In practice, it is common to work with multiple time series that are inter-related and non-stationary. We propose a method to normalize univariate time series conditional on a set of covariates. This method can be considered as a variation of z-score normalization, but where the mean and variance are functions of the covariates. Thus we refer to this method as \emph{conditional normalization}.

In the proposed method, we first estimate the conditional mean of the time series using a generalized additive model (GAM) \autocite{hastie1990generalized} with a set of covariates. The conditional variance is then estimated via a different GAM fitted to the squared errors from the conditional mean model, with respect to the same set of covariates. Finally, the estimated conditional mean and variance are used to standardize the time series. One can choose the most relevant set of covariates that can explain maximum variation in the time series.

It is relatively common to subtract a conditional mean in order to adjust data for subsequent analysis, and sometimes this is called ``normalization'' \autocite[e.g.,][]{xie2019}. However, our approach is much more general as both the conditional mean and conditional variance are modeled, and we allow for non-linear relationships between the response and covariates in both models.

We show two possible uses of conditionally normalized time series. First, we describe how the conditionally normalized time series can be used to impute missing values in a univariate time series. To do this, we model the normalized series, and use the model to impute the missing values. The resulting imputations are then ``unnormalized'' to give estimates on the original scale.

Second, we show how the conditionally normalized time series can be used to estimate the conditional Autocorrelation Function (ACF) and conditional Cross-Correlation Function (CCF). We can define the conditional ACF at lag \(k\) as the \emph{conditional expectation of the cross-product of the conditionally normalized time series and its \(k^{\text{th}}\) lagged series}. Similarly, the conditional CCF at lag \(k\) can be defined as the \emph{conditional expectation of the cross-product between two conditionally normalized time series at \(k\) lags apart}. To estimate the conditional expectations, we propose to fit GAMs for the cross-product of the normalized time series using the same set of covariates used in the conditional normalization.

We also highlight two straightforward empirical applications of conditional normalization of time series. The first application involves a time series of mean daily stream temperatures observed in multiple locations in Boise River, in the northwestern United States of America (USA), which has many missing values. We describe how we can impute these missing values using Bayesian machinery for modeling.

The second application uses the conditional CCF to estimate the lag time between two sensor locations in the Pringle Creek river network in Texas, USA. The lag time is the time it takes water to flow downstream from an upstream location. This lag time often depends on the upstream river behavior. For example, when the upstream water level increases, water flow will typically be increased and hence the lag time will decline. On the other hand, when the level is low, water may be flowing more slowly and hence the lag time will increase. Lag time has been estimated using different approaches in many hydrological applications (see \textcite{van2010}, \textcite{hrachowitz2016} and \textcite{li2018} for example). We propose to estimate the lag time as the lag that gives the maximum conditional cross-correlation between two water-quality variables observed at upstream and downstream locations, conditional on other, related water-quality variables measured at an upstream location. This will allow the lag time to be estimated conditional on the upstream river behavior.

The rest of the paper is organized as follows. The underlying methods for conditional normalization are described in \autoref{sec:methodology}. \autoref{sec:app1} contain the empirical application on missing value imputation, while \autoref{sec:app2} discusses the application to estimating lag times between sensor locations. Finally, we discuss the results and provide some concluding remarks in \autoref{sec:discussion}.

\hypertarget{sec:methodology}{%
\section{Conditional estimation via GAMs}\label{sec:methodology}}

In this section, we discuss our approach to conditional normalization of a time series, and then we discuss a couple of scenarios in which these conditionally normalized time series can be helpful in data analysis.

\hypertarget{sec:cond-norm}{%
\subsection{Conditional normalization}\label{sec:cond-norm}}

Let \(y_t\) be a variable observed at times \(t = 1,\dots,T\), and \(\bm{z}_t = (z_{1,t},\dots,z_{p,t})\) be a \(p\) dimensional vector of variables measured at the same times. We assume the mean and variance of \(y_t\) are functions of \(\bm{z}_t\); that is, \(\text{E}(y_t\mid\bm{z}_t) = m(\bm{z}_t)\) and \(\text{V}(y_t\mid\bm{z}_t) = v(\bm{z}_t)\). Our aim is to normalize \(y_t\) conditional on \(\bm{z}_t\), giving
\begin{equation}\label{eq:cond_norm}
y^*_{t} = \frac{y_{t} - \hat{m}(\bm{z}_t)}{\sqrt{\hat{v}(\bm{z}_{t})}}.
\end{equation}

We estimate \(m(\bm{z}_t)\) and \(v(\bm{z}_t)\) using GAMs \autocite{hastie1990generalized}. First, we fit the model
\[
  y_t = \alpha_0 + \sum_{i=1}^p f_{i}(z_{i,t}) + \varepsilon_{t},
\]
where \(f_{i}(\cdot)\) are smooth functions, and \(\varepsilon_{1},\dots,\varepsilon_T\) have mean 0 and variance \(v(\bm{z}_t)\), giving
\begin{equation}\label{eq:cond_mean}
\hat{m}(\bm{z}_t) = \hat{\alpha}_0 + \sum_{i=1}^p \hat{f}_{i}(z_{i,t}).
\end{equation}
In estimating the model, we ignore any heteroskedasticity and autocorrelation, and use penalized splines for each \(f_i\) function.

Next, we fit the model
\begin{align*}
[y_t- \hat{m}(\bm{z}_t)]^2 & \sim \text{Gamma}(v(\bm{z}_t), r),\\
  \log(v(\bm{z}_t)) &= \beta_0 + \sum_{i=1}^p g_{i}(z_{i,t}),
\end{align*}
where each \(g_i(\cdot)\) is a smooth function, and the Gamma parameterization has the first argument, \(v(\bm{z}_t)\), as the mean, and the second argument, \(r\), as the shape parameter. The Gamma family is not essential here, and it may be replaced by another distribution whose support is in \((0, \infty)\). The resulting variance estimate is
\begin{equation}\label{eq:cond_var}
\hat{v}(\bm{z}_t) = \exp\bigg(\hat{\beta}_0 + \sum_{i=1}^p \hat{g}_{i}(z_{i,t})\bigg).
\end{equation}

\hypertarget{sec:imputing_missing}{%
\subsection{Imputation of missing values}\label{sec:imputing_missing}}

The conditionally normalized series can be used when imputing missing values in a univariate time series, assuming \(y_t^*\) is a (possibly seasonal) Autoregressive Integrated Moving Average (ARIMA) process. The normalization removes some of the sources of variation in the data, allowing the ARIMA model to handle any remaining serial correlation.

We can then impute \(y_t\) using
\begin{equation}\label{eq:univ_interpolate}
y_t = \hat{y}_t^*\sqrt{\hat{v}(\bm{z}_t)} + \hat{m}(\bm{z}_t).
\end{equation}
where \(\hat{y}_t^*\) is the imputed value of \(y_t^*\), computed using a Kalman smoother.

\hypertarget{conditional-autocorrelation-function}{%
\subsection{Conditional autocorrelation function}\label{conditional-autocorrelation-function}}

We can also use the normalized time series to compute a conditional ACF as
\[
  r_k(\bm{z}_t) = \text{E}[y_t^*y^*_{t-k}\mid\bm{z}_t] \quad \text{for} \quad k = 1,2,\dots
\]
The function \(r_k(\cdot)\) can be estimated using a separate GAM for each \(k\):
\begin{align*}
y_t^*y_{t-k}^* \sim N(r_k(\bm{z}_t), \sigma_k^2), \\
\eta(r_k(\bm{z}_t)) = \gamma_0 + \sum_{i=1}^p h_i(z_{i,t}),
\end{align*}
where \(h_i(.)\) are smooth functions and
\begin{equation}\label{eq:inverse_link}
  \eta^{-1}(u) = \frac{e^u - 1}{e^u + 1}.
\end{equation}
Other smooth monotonic link functions, \(\eta\), that map \([-1,1]\) to the real line, \((-\infty, \infty)\), may also be used.

\hypertarget{sec:cross-correlation}{%
\subsection{Conditional cross-correlation function}\label{sec:cross-correlation}}

We can use a similar approach to estimate the conditional cross-correlation functions. Suppose we have a variable \(x_t\) observed at the same time as \(y_t\). We are interested in estimating the cross-correlation between \(x_t\) and \(y_{t+k}\) for \(k = 1,2, \dots\), conditional on a set of variables \(\bm{z}_t\). First we normalize \(x_t\) and \(y_{t}\) with respect to \(\bm{z}_t\) using \eqref{eq:cond_norm} and get
\[
  x^*_t = \frac{x_{t} - \hat{m}_x(\bm{z}_t)}{\sqrt{\hat{v}_x(\bm{z}_{t})}}
    \qquad\text{and}\qquad
  y^*_{t} = \frac{y_{t} - \hat{m}_{y}(\bm{z}_t)}{\sqrt{\hat{v}_{y}(\bm{z}_{t})}},
\]
where \(m_x(z_t)\) and \(m_{y}(z_t)\) are estimated using \eqref{eq:cond_mean}, and \(v_x(z_t)\) and \(v_{y}(z_t)\) are estimated using \eqref{eq:cond_var}.

Then we can estimate the conditional cross-correlation
\[
  c_k(\bm{z}_t) = \text{E}[y_{t+k}^*x^*_t\mid\bm{z}_t] \quad \text{for} \quad k = 1,2,\dots
\]
using the GAMs
\begin{align*}
y_{t+k}^*x_t^* \sim N(c_k(\bm{z}_t), u_k^2),
\end{align*}
\begin{equation}\label{eq:ccf_gam}
\eta(c_k(\bm{z}_t)) = \phi_0 + \sum_{i=1}^p s_i(z_{i,t}).
\end{equation}

\hypertarget{sec:app1}{%
\section{Application: Stream temperature imputation}\label{sec:app1}}

\hypertarget{temperature-data}{%
\subsection{Temperature data}\label{temperature-data}}

In this case study, we use a dataset comprising daily mean stream temperatures (\(^\circ\)C) recorded in a large-scale dendritic network in northwestern USA \autocite{isaak2017norwest}. The five-year dataset includes data from 42 in-situ sensors, each deployed in a unique spatial location. It contains mean daily stream temperature data, with a total of 1825 observations per time series. There is a tendency to get missing values in the original data due to sensor issues and the goal is to impute those values. For illustration purposes, we took a subset of the data at evenly spaced intervals discarding all but every 5th observation in the time series resulting in 73 observations per year on each spatial location.

\begin{figure}
\includegraphics[width=1\linewidth]{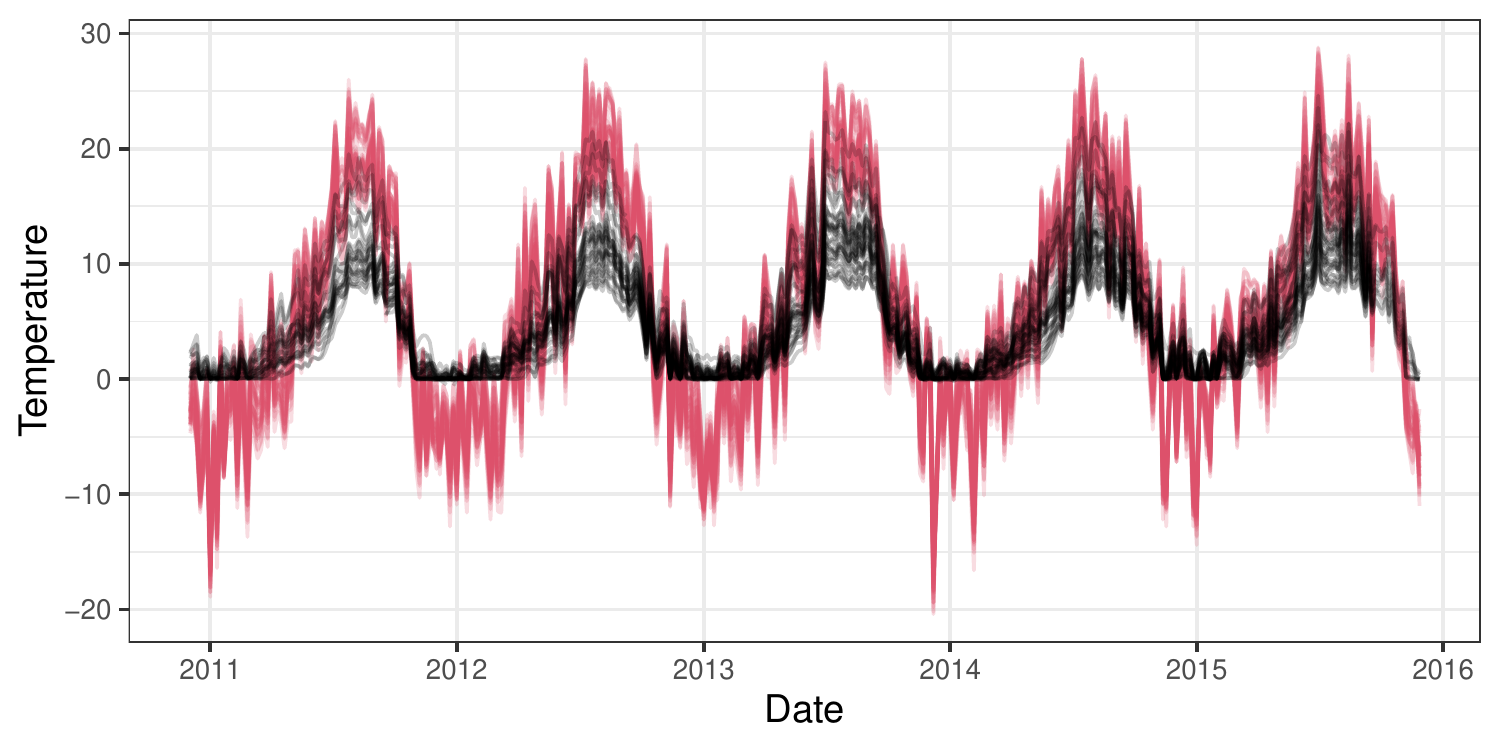} \caption{Stream temperatures (black) and air temperatures (red) ($^\circ$C) from the 42 spatial locations.}\label{fig:timeSeriesCode}
\end{figure}

\autoref{fig:timeSeriesCode} shows the time series of stream temperature and air temperature. It is known that the air temperatures are strongly correlated with stream temperatures \autocite{bal2014hierarchical}, and so these will be used as a covariate in our models.

\hypertarget{the-conditional-normalization-model}{%
\subsection{The conditional normalization model}\label{the-conditional-normalization-model}}

Let \(y_{st}\) represent the temperature at spatial locations \(s = 1,2,\dots,S\), and time points \(t = 1,2,\dots,T\), where \(S=42\) and \(T=1825\). We will use the conditional normalization model:
\[
  y_{st}^* = \frac{(y_{st} - \mu_{st})}{\sigma_{t}} \sim \text{AR}(p),
\]
where \(\mu_{st}\) and \(\sigma_{t}\) are the mean and the standard deviation respectively (formulated as functions of covariates). To avoid overparameterization (i.e.~having more parameters than what we can learn from the model), we assume that sites have a common standard deviation \(\sigma\) at time \(t\). The standardized response variable \(y_{st}^*\) is modeled using an autoregressive process of order \(p\) to account for the remaining serial correlation in the data.

\hypertarget{gam-models}{%
\subsection{GAM models}\label{gam-models}}

We formulate the expected value (\(\mu_{st}\)) of stream temperature as a function of air temperatures, as well as spatial covariates and landscape factors that are constant over time: stream slope, elevation (elev) and cumulative drainage area (cd). The resulting mean is the linear function
\begin{align*}
\mu_{st} = \beta_0
          & + \beta_1 \textrm{slope}_{s}
            + \beta_2 \textrm{elev}_{s}
            + \beta_3 \textrm{cd}_{s}
            + \beta_4 \textrm{at}_{st}  \\
          & + \beta_5 \sin_{1t}
            + \beta_6 \cos_{1t}
            + \beta_7 \sin_{2t}
            + \beta_8 \cos_{2t}
            + \beta_9 \sin_{3t}
            + \beta_{10} \cos_{3t} \\
          & + \beta_{11} \sin_{4t}
            + \beta_{12} \cos_{4t}
            + \beta_{13} \sin_{5t}
            + \beta_{14} \cos_{5t}.
\end{align*}
The covariates \(\sin_{kt} = \sin(2\pi tk/m)\) and \(\cos_{kt}= \cos(2\pi tk/m)\), \(k=1,\dots,5\), are the first five pairs of Fourier terms (harmonic regression parameters), where \(m\) is the seasonal period \autocite[Section 7.4,][]{fpp3}.

We model \(\sigma_t\) using a gamma distribution with common \(a\) and time-specific parameter \(b_t\), \(\sigma_{t}^2 \sim \text{Gamma}(a, b_t)\). We use a non-informative uniform prior for \(a \sim U(0, 100)\) and set \(b_t = a / \exp{(\bm{X} \bm{\gamma})}\), where \(\bm{X}\) is a design matrix containing the Fourier covariates and \(\bm{\gamma}\) is a vector of regression coefficients.

\hypertarget{results}{%
\subsection{Results}\label{results}}

\begin{figure}
\includegraphics[width=1\linewidth]{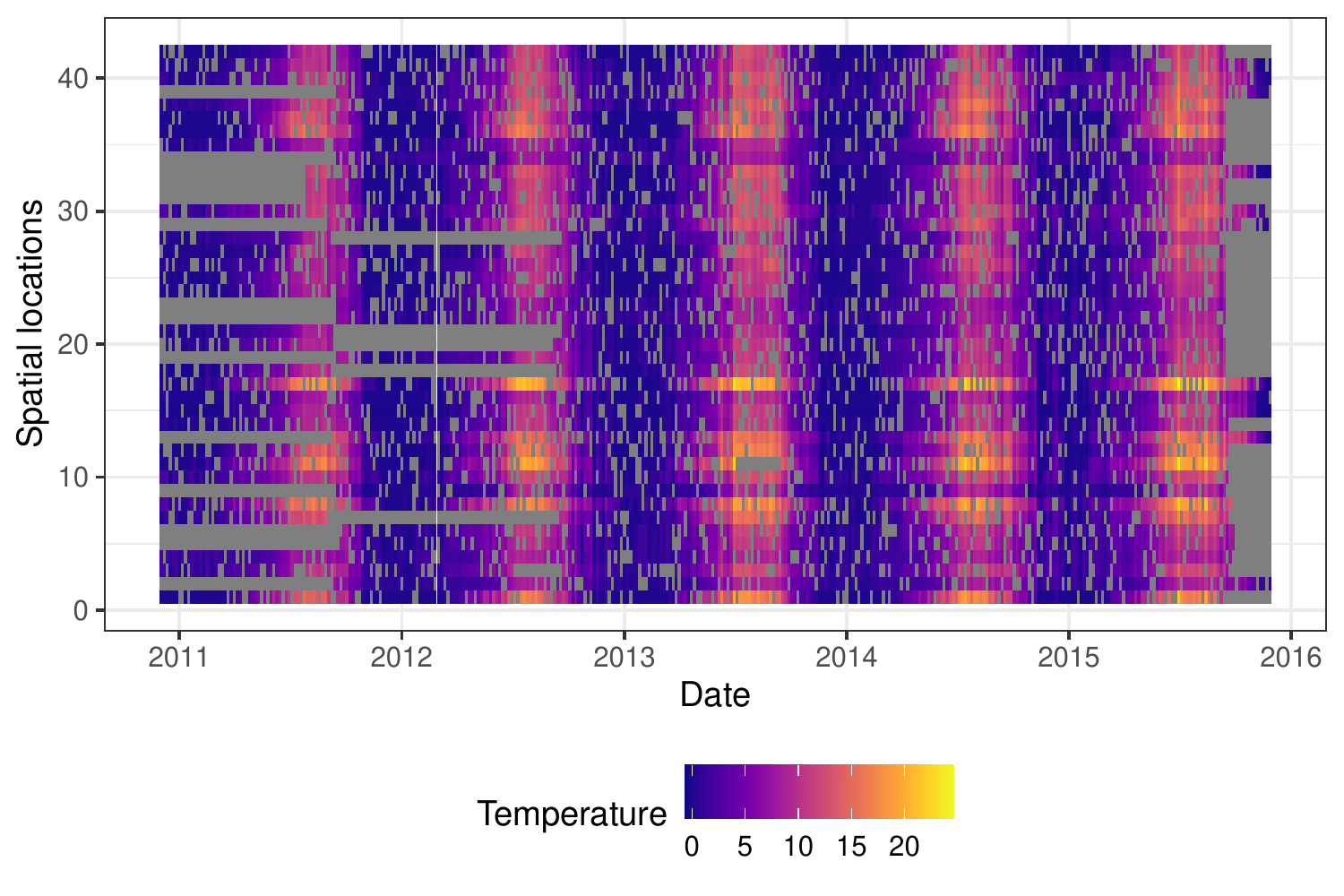} \caption{Daily mean temperature ($^\circ$C) values. The gray areas represent periods that are missing from the training data.}\label{fig:train-testing}
\end{figure}

\begin{figure}
\includegraphics[width=1\linewidth]{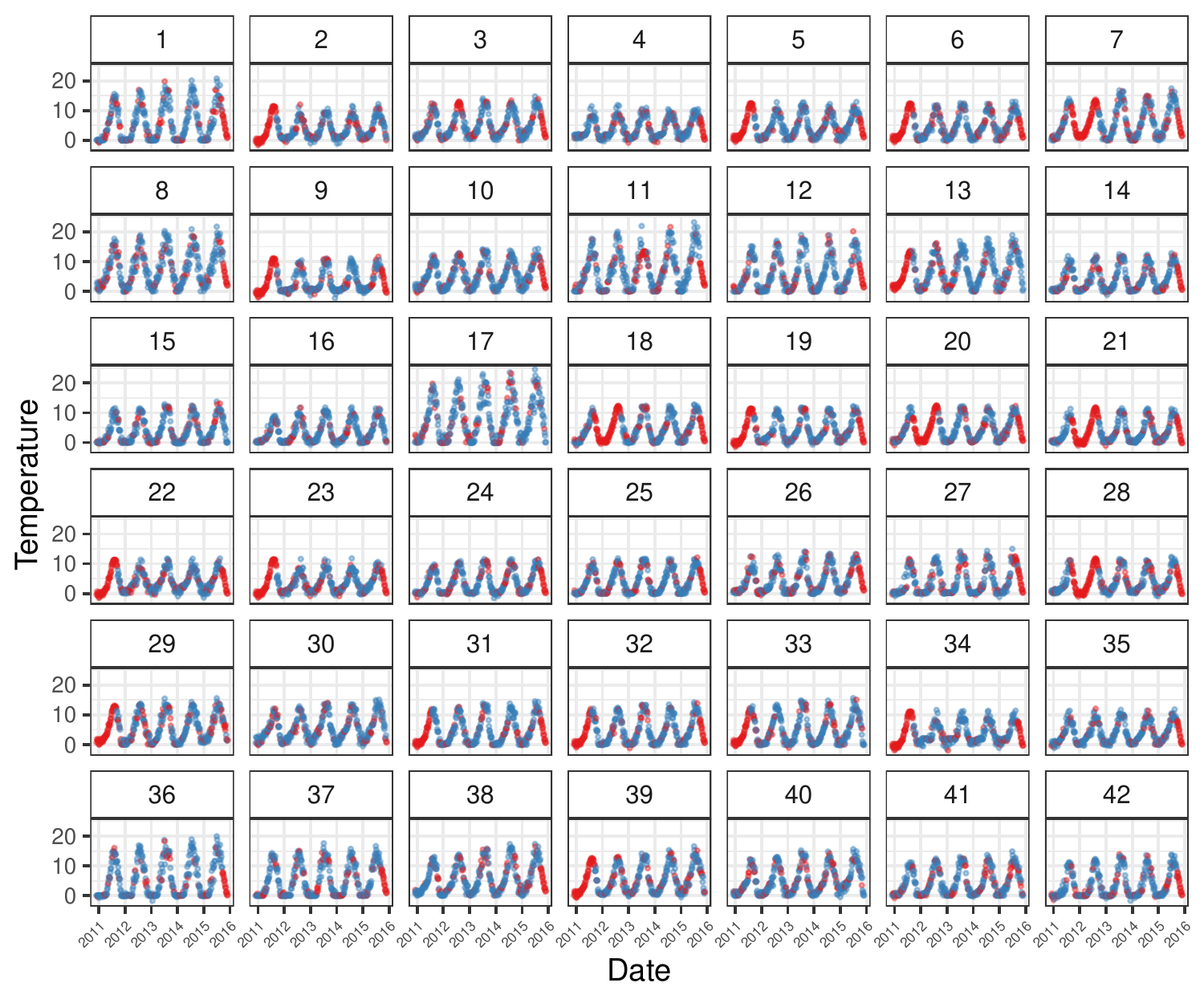} \caption{Time series of stream temperature in the 42 spatial locations. Points in blue represent the training set, while the predictions for the missing periods are given in red.}\label{fig:preds}
\end{figure}

\begin{figure}
\includegraphics[width=1\linewidth]{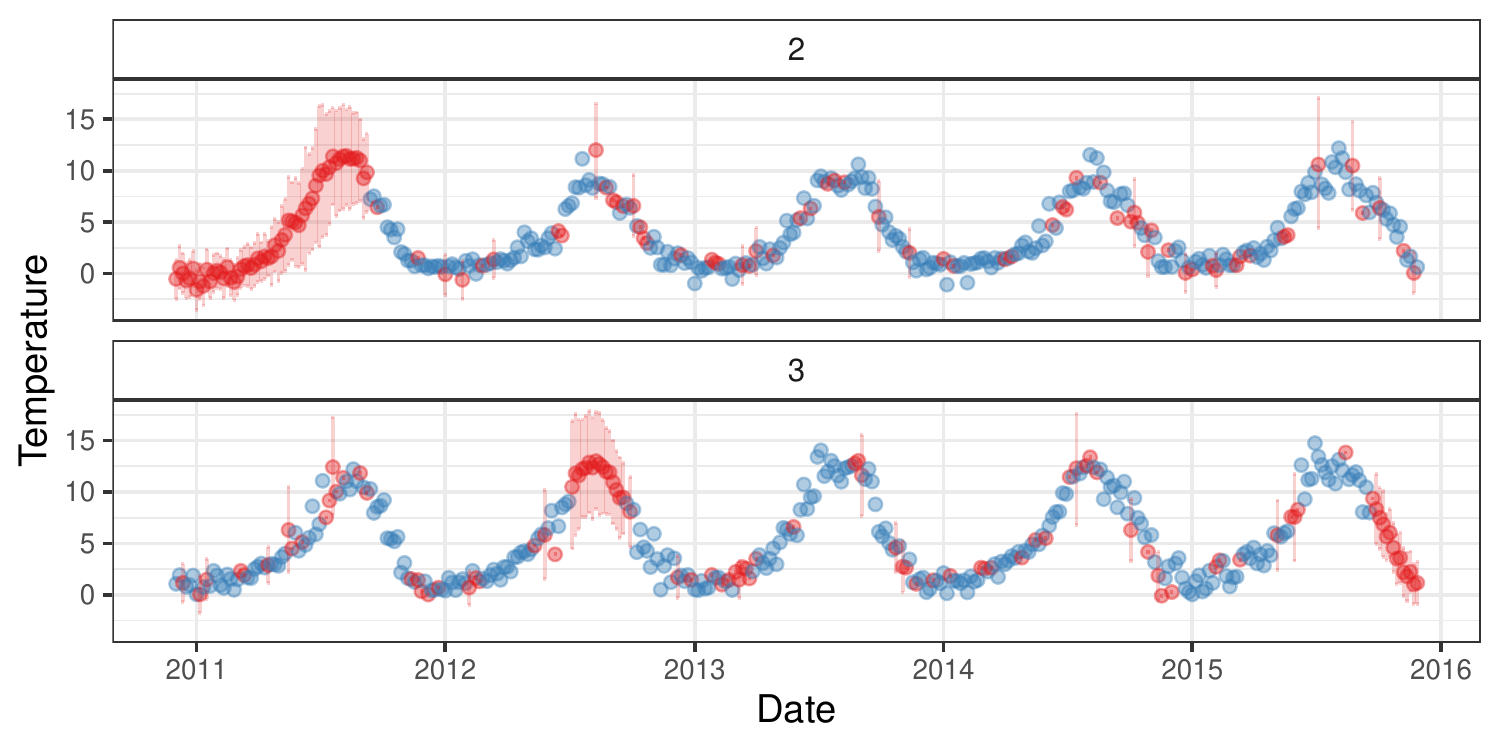} \caption{Time series of two spatial locations. Points in blue represent the training set, while the predictions for the missing periods are given in red along with the 95\% highest posterior density intervals.}\label{fig:unc}
\end{figure}

The data are missing 1654 temperature values out of 15330 observation periods. To illustrate our approach, we also remove 20\% of the non-missing observations to form a test set. We aim to estimate these missing values and compare the estimates with the original values. Thus, the model is trained using 80\% of the non-missing data, with 20\% used for testing the prediction accuracy. \autoref{fig:train-testing} shows the water temperature values in the training data for each of the spatial locations.

The model is estimated in Stan using a Hamiltonian Monte Carlo procedure. We use 3 chains each composed of 6,000 samples and we discard a burn-in of 4,000 samples.

We found that the AR of order eight offered the best fit in terms of root mean square prediction error (RMSPE). \autoref{fig:preds} shows the observed water temperature (training set = blue) and the estimated values (testing set = red) in each of the 42 spatial locations. We note that the model captures well the periodicity in the data and produces good estimates of the missing temperature values when comparing the predictions vs hold-out data.

We also assessed the uncertainty in the estimates using highest posterior density intervals. \autoref{fig:unc} shows these for the first two spatial locations.
In both locations, the model captures the periodic patterns, including backcasting to predict the initial missing sections of the time series in location two. Overall, the proportion of observations in which the nominal 95\% highest density interval of the estimated mean temperature contains the true value is 0.946.

\begin{figure}
\includegraphics[width=1\linewidth]{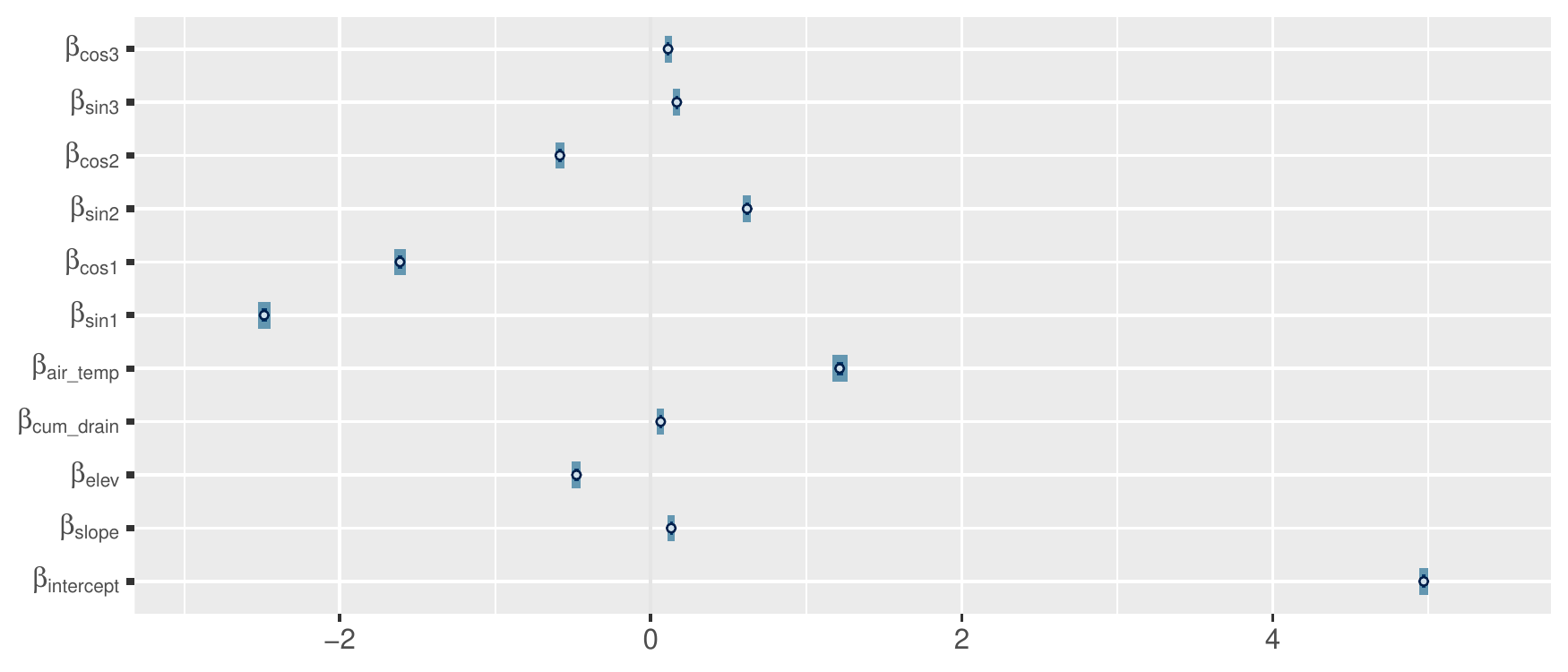} \caption{Posterior means of the regression coefficients.}\label{fig:reg-coef}
\end{figure}

The posterior distributions of the regression coefficients \(\bm{\beta}\) indicate that the spatial covariates and air temperature substantially affect the stream temperature (\autoref{fig:reg-coef}), with the three pairs of Fourier terms explaining seasonality changes in the response variable well.

\begin{figure}
\includegraphics[width=1\linewidth]{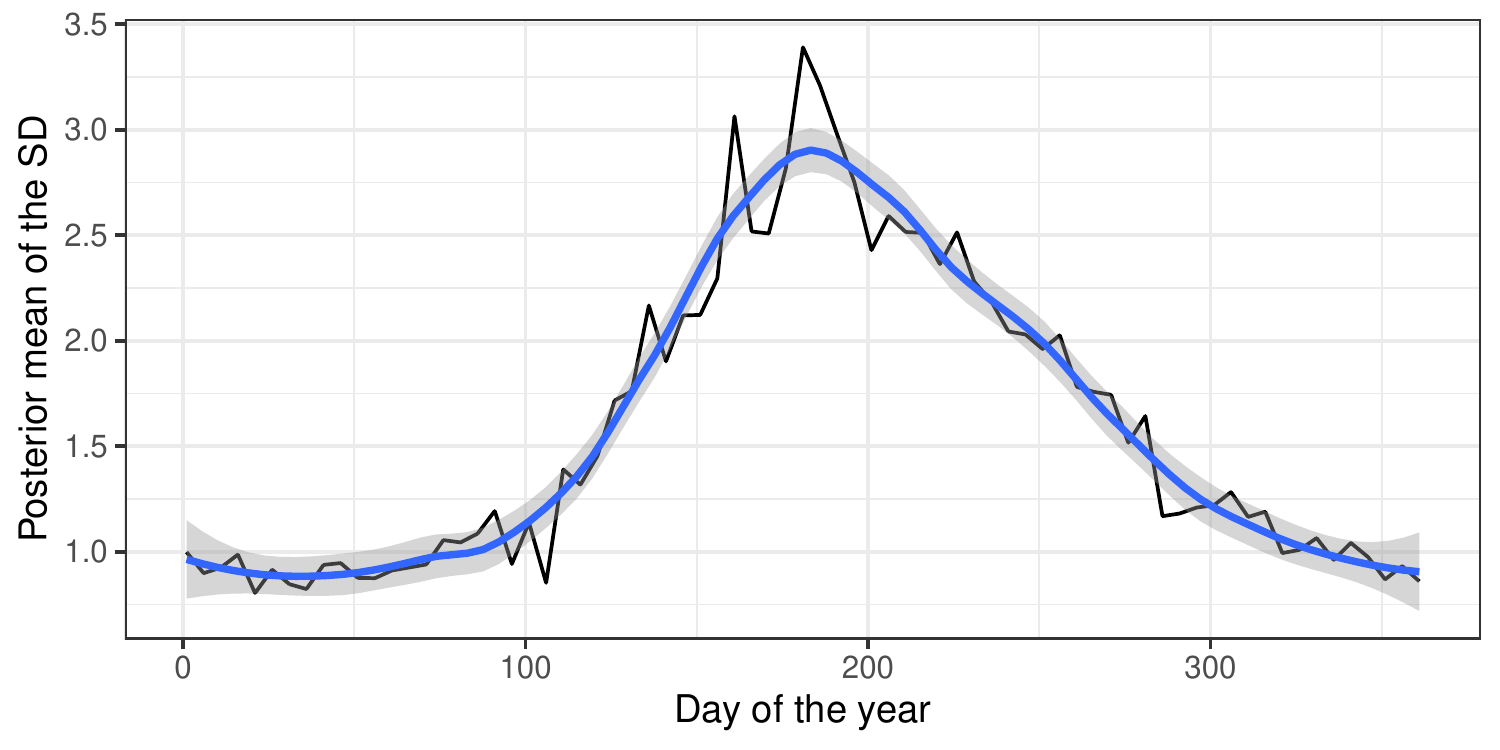} \caption{Posterior means of the standard deviation ($\sigma$).}\label{fig:sd}
\end{figure}

The posterior means of the daily standard deviation indicates that the standard deviation in summer was three times higher than in the winter (\autoref{fig:sd}).

An ACF plot of the standardized time series corresponding to site 2 is presented in \autoref{app:appendixA}.

\hypertarget{sec:app2}{%
\section{Application: Predicting lag time on river flow}\label{sec:app2}}

\hypertarget{automated-in-situ-sensors}{%
\subsection{Automated in-situ sensors}\label{automated-in-situ-sensors}}

Traditional methods of monitoring water quality include collecting water samples at low frequencies (monthly, bi-monthly) and conducting lab-based assessments to measure water quality. With advancements in technology, this manual process is being replaced or augmented by automated, high-frequency in-situ sensors.

In river systems, these in-situ sensors are typically placed at one or more locations to measure multiple water-quality variables semi-continuously. The resultant data can be used for different kinds of analyses such as identifying trends in water quality and predicting sediment and nutrient concentrations through space and time \autocite{leigh2019}.

\hypertarget{lag-time}{%
\subsection{Lag time}\label{lag-time}}

When analysing sensor data from multiple locations along the same water flow path, it is useful to know the lag time between sensor locations, as this facilitates prediction of downstream water quality using information collected upstream. Lag time can be defined hydrologically in many ways. For example, \textcite{wanielista1997} defined it as
``\emph{time from the centroid of rainfall excess to the time of peak runoff for a watershed}''. Here we define the lag time specifically as ``\textbf{the time it takes water to flow downstream from an upstream location}''.

Estimating lag time is necessary in order to compute lagged, explanatory water-quality variables which can be used as predictors in models of the downstream response variable of interest. One approach for estimating lag time is to use empirical equations based on the length and slope of the flow path and other catchment features \autocite{Green2002,li2008}. Another approach uses water level and flow (i.e., discharge) data. For example, \textcite{seyam2014} used this approach to estimate lag time between four upstream locations and a downstream location in the Selangor River basin. Their method involved plotting the hydrograph for the downstream location and then water level at the upstream location during high flow events and estimating the time difference between peaks of the two plots. The average time difference was then considered as the lag time between the two locations.

Field-based methods to estimate the lag time include injecting salt tracers (usually Sodium Chloride or Sodium Bromide) at an upstream location and measuring the salt concentration through time at a downstream location, from which the travel time is then estimated. This manual process has to be carried out several times a year during both high flows and low flows, which is costly and time-consuming.

\hypertarget{estimating-lag-time-via-conditional-cross-correlations}{%
\subsection{Estimating lag time via conditional cross-correlations}\label{estimating-lag-time-via-conditional-cross-correlations}}

Lag time can be influenced by various environmental conditions upstream, such as the water level, discharge, temperature and other water-quality variables. Therefore, we propose a method to estimate the lag time between two sensor locations in a river using conditional cross-correlations.

Suppose \(x_t\) and \(y_t\), observed at times \(t=1,\dots,T\), denote the same water-quality variable measured at upstream and downstream sensors respectively. Let \(\bm{z}_t\) be a \(p\) dimensional vector of other water-quality variables measured at the upstream sensor at time \(t\). We estimate \(c_k(\bm{z}_t)\), the cross-correlation between \(y_{t+k}\) and \(x_t\) at lags \(k = 1,\dots,K\), conditional on \(\bm{z}_t\), using the model described in \autoref{sec:cross-correlation}. Then, we can use the estimates of conditional cross-correlations to estimate the river lag time between the two locations. We define this lag time as \(d_t\), and estimate it using
\begin{equation}\label{eq:dt}
\hat{d}_{t}(\bm{z}_t) = \underset{k}{\operatorname{argmax}}~ \hat{c}_{k}(\bm{z}_t).
\end{equation}

\hypertarget{sec:SieveBSalgo}{%
\subsection{\texorpdfstring{Computing bootstrapped confidence intervals for \(d_t\)}{Computing bootstrapped confidence intervals for d\_t}}\label{sec:SieveBSalgo}}

Computing the standard errors and confidence intervals for \(d_t\) is not straightforward, so we use a bootstrap method. We resample the residuals from the various models used in the conditional cross-correlation calculation to generate new data. Because these residuals are serially correlated, we use a sieve bootstrap approach \autocite{Buhlmann1997} to capture the autocorrelation structure in the data in our bootstrap samples. The following algorithm describes our approach for computing these confidence intervals.

\hypertarget{algorithm-sieve-bootstrap-confidence-intervals-for-d_t}{%
\subsubsection*{\texorpdfstring{Algorithm (Sieve bootstrap confidence intervals for \(d_t\))}{Algorithm (Sieve bootstrap confidence intervals for d\_t)}}\label{algorithm-sieve-bootstrap-confidence-intervals-for-d_t}}
\addcontentsline{toc}{subsubsection}{Algorithm (Sieve bootstrap confidence intervals for \(d_t\))}

Recall that we have fitted the following separate GAMs for each \(k\)
\[
y^*_tx^*_{t-k} = \eta^{-1}(\phi_0 + \sum_{i=1}^{p_k}s_i(z_{i,t})) + \varepsilon_{t,k}.
\]
Since the \(\varepsilon_{t,k}\) are serially correlated, we fit a \(p_k^{\text{th}}\) order autoregressive model for \(\varepsilon_{t,k}\) for each \(k\):
\[
\varepsilon_{t,k} = \mu_k + \sum_{i=1}^{p_k}\psi_{i,k}\varepsilon_{t-i,k} + \zeta_{t,k}, \quad \zeta_{t,k} \sim N(0,\sigma_k^2).
\]
For each model, the order \(p_k\) is determined by minimizing the corrected Akaike Information Criterion (AICc) using the \texttt{auto.arima} function from the \texttt{forecast} package \autocite{forecast,HK08}. Then we resample from \(\hat{\zeta}_{t,k}\) to generate our bootstrap sample following these steps.

\begin{enumerate}
\def\labelenumi{\arabic{enumi}.}
\tightlist
\item
  Randomly select with replacement a sample of size \(T\) from \(\hat{\zeta}_{t,k} = \varepsilon_{t,k} - \hat{\mu}_k - \sum\limits_{i=1}^{p_k}\hat{\psi}_{i,k}\varepsilon_{t-i,k}\). Denoted this sample as \(\hat{\zeta}_{t,k}^b\).
\item
  Compute \(\varepsilon_{t,k}^b\) as \(\varepsilon_{t,k}^b = \hat{\mu}_k + \sum\limits_{i=1}^{p_k}\hat{\psi}_{i,k}\varepsilon^b_{t-i,k} + \zeta_{t,k}^b\) for \(k = 1,\dots,K\).
\item
  Compute \({y^*_tx^*_{t-k}}^b = \eta^{-1}(\hat{\phi}_0 + \sum\limits_{i=1}^{p_k}\hat{s}_i(z_{i,t})) + \varepsilon_{t,k}^b\) for \(k = 1,\dots,K\).
\item
  Fit the following GAM to the bootstrapped data for each \(k\):
  \[
  {y^*_tx^*_{t-k}}^b = \eta^{-1}(\phi_0 + \sum_{i=1}^{p_k}s_i(z_{i,t})) + \varepsilon_{t,k}.
  \]
\item
  Use the models in step 4 to compute \(d_t\) for a given set of \(z_i\).
\item
  Repeat steps 1 to 5 for \(b=1,\dots,m\), where \(m=1000\). Thereby we will get a sample of \(d_t\) of size \(m\) which can be used to form the empirical distribution of \(d_t\). Use this sample to compute \((\alpha/2)^{\text{th}}\) and \((1 - \alpha/2)^{\text{th}}\) quantiles which represent the lower and upper bounds for the \(100(1-\alpha)\%\) confidence interval for \(d_t\).
\end{enumerate}

\hypertarget{study-area-and-water-quality-data}{%
\subsection{Study area and water-quality data}\label{study-area-and-water-quality-data}}

We consider Pringle Creek, one of the NEON (National Ecological Observatory Network) aquatic sites located in Wise County, Texas, and managed by the U.S Forest Services\footnote{\url{https://www.neonscience.org/field-sites/prin}}. A detailed description of the study site is given in \autoref{app:appendixB}.

Water quality is measured in Pringle Creek using two sensor locations situated about 200~m apart, with a small tributary entering the main creek between the two sensors. The variables measured by these sensors include turbidity (Formazin Nephelometric Unit), specific conductance, pH, dissolved oxygen, and chlorophyll. Measurements are available at 1-minute frequencies and can be retrieved from NEON Data Portal \autocite{NEON_data_WQ}. Surface water level and water temperature are also available from both locations at 5-minute frequencies and can be retrieved from \textcite{NEON_data_level} and \textcite{NEON_data_temp}, respectively.

The data we consider were collected from 1~October 2019 to 31~December 2019. This time span avoids the summer period in which surface pools of water disconnect and contains the least number of missing observations after removing the anomalies.

We will use turbidity to compute the cross-correlations between upstream and downstream sensors. Turbidity is chosen because it is heavily influenced by fresh inputs of water from upstream, and hence there should be a strong relationship between upstream and downstream turbidity. We choose water level and temperature as the covariates from the upstream sensor to model the cross-correlation between upstream and downstream turbidity.

\autoref{app:appendixB} discusses the data pre-processing, anomaly detection, missing value imputation, and variable selection steps of the analysis.

\hypertarget{conditional-cross-correlation-between-turbidity-at-upstream-and-downstream-sensors}{%
\subsection{Conditional cross-correlation between turbidity at upstream and downstream sensors}\label{conditional-cross-correlation-between-turbidity-at-upstream-and-downstream-sensors}}

Based on the NEON reaeration sampling protocol \autocite{Neon_reaeration} and information gathered from field experts, the time it takes water to travel downstream between two sensor locations at Pringle Creek is typically about \(45-60\) minutes, though it may be shorter than \(45\) minutes during high flows and greater than \(60\) minutes during low flows. Considering this information, we choose to compute the cross-correlations up to \(24\) lags, which will allow for a maximum of two hours of travel time (as the frequency of the water-quality data is \(5\) minutes).

\begin{figure}
\includegraphics[width=1\linewidth]{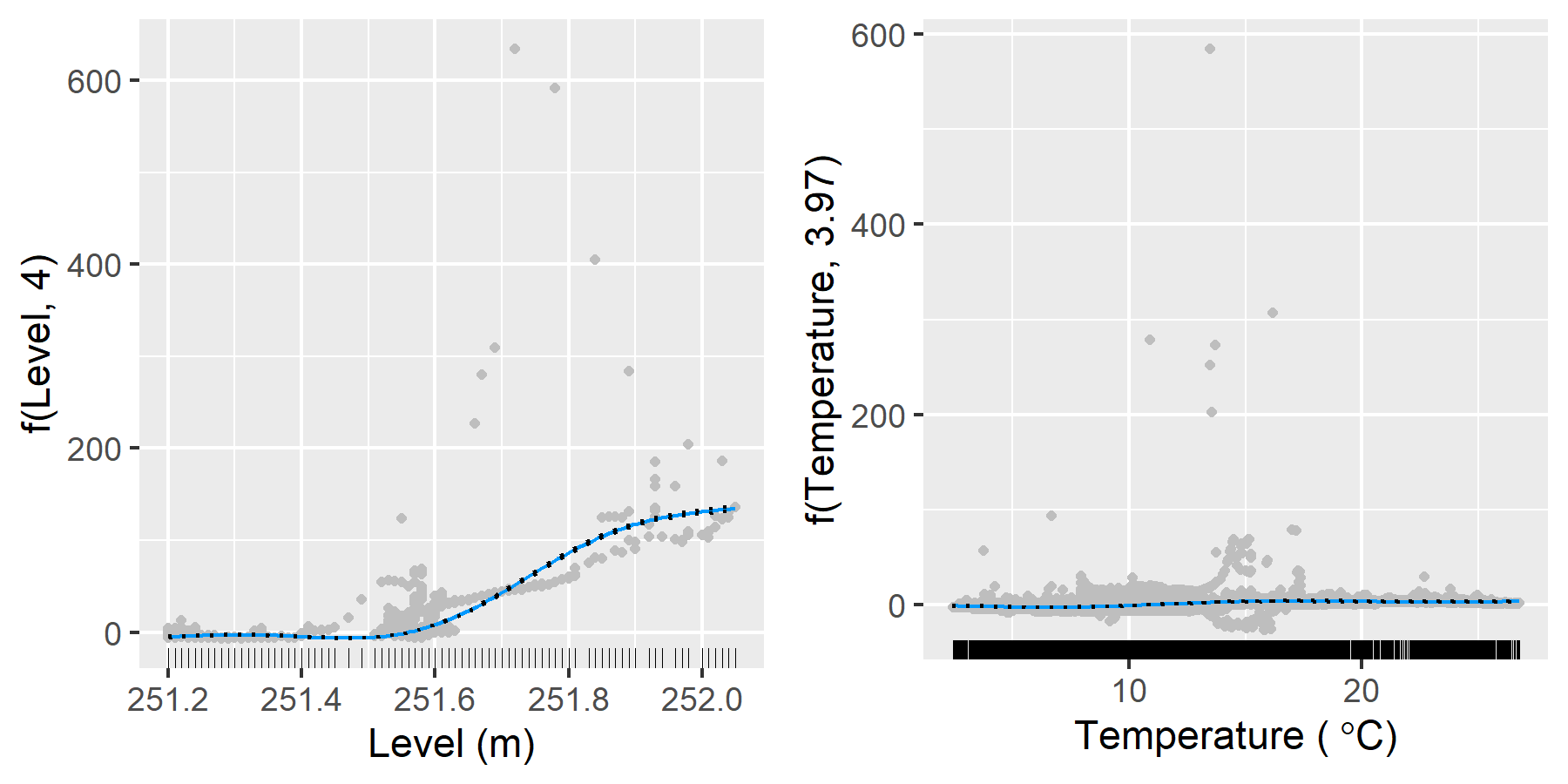} \caption{Visualizing the fitted smooth functions in the conditional mean model for turbidity downstream with the predictors, water level and temperature from upstream sensor. Each panel visualizes the relationship between the response and predictor while holding other predictors at their medians ($251.6m$ and $9.926^{o}C$ for water level and temperature, respectively). The smooth function is shown in blue with 95\% confidence bands. The degrees of the smoothing are shown in the y-axis label for each plot.}\label{fig:cond-mean-turbidity-downstream}
\end{figure}

\begin{figure}
\includegraphics[width=1\linewidth]{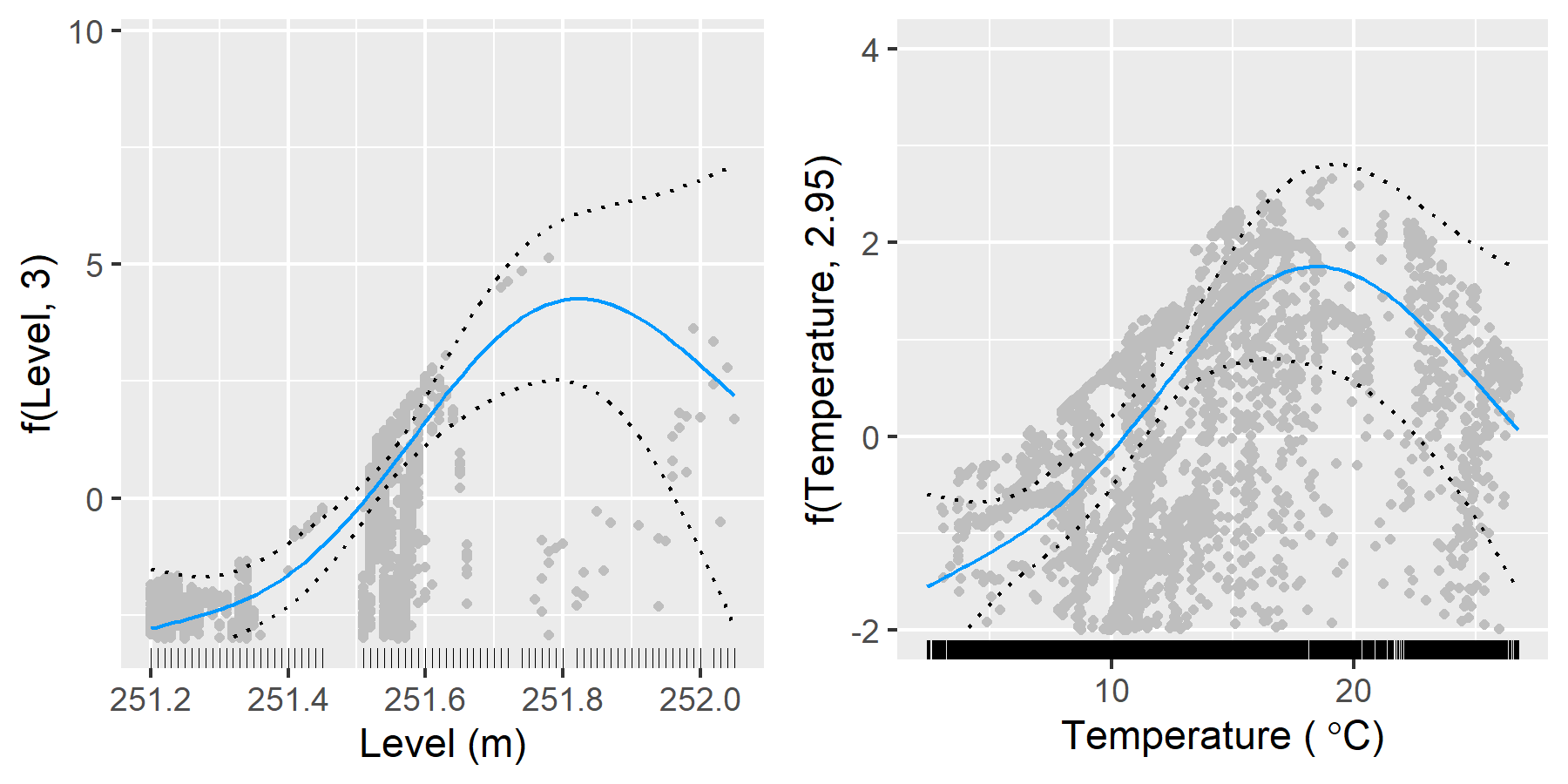} \caption{Visualizing the fitted smooth functions in the conditional variance model for turbidity downstream with the predictors, water level and temperature from upstream sensor. Each panel visualizes the relationship between the response and predictor while holding other predictors at their medians ($251.6m$ and $9.926^{o}C$ for water level and temperature, respectively). The smooth function is shown in blue with 95\% confidence bands. The degrees of the smoothing are shown in the y-axis label for each plot.}\label{fig:cond-var-turbidity-downstream}
\end{figure}

Let \(y_t\) denote the time series of turbidity measured at the downstream sensor. Following \autoref{sec:cross-correlation}, we first normalize \(x_t\) and \(y_{t+k}\) for \(k = 1,\dots, 24\) conditional on \(\bm{z}_t\). Figures \ref{fig:cond-mean-turbidity-downstream} and \ref{fig:cond-var-turbidity-downstream} visualized the fitted mean and variance models for \(y_{t+1}\) respectively.

Following Equation \eqref{eq:ccf_gam}, the fitted conditional cross-correlation function between turbidity at the upstream and downstream sensors at lag \(k\) can be written as
\begin{equation}\label{eq:ccf_k}
  \hat{c}_k(\bm{z}_t) = \eta^{-1}\big(\hat{\phi}_o + \hat{s}_{1,k}(\text{level\_upstream})_t + \hat{s}_{2,k}(\text{temperature\_upstream})_t),
\end{equation}
where \(\{\hat{s}_{1,k}, \hat{s}_{2,k}\}\) denote natural cubic splines. Similar to when fitting mean and variance models, the degrees of freedom for each spline are chosen by examining the relationship between the response and each covariate.

\begin{figure}
\includegraphics[width=1\linewidth]{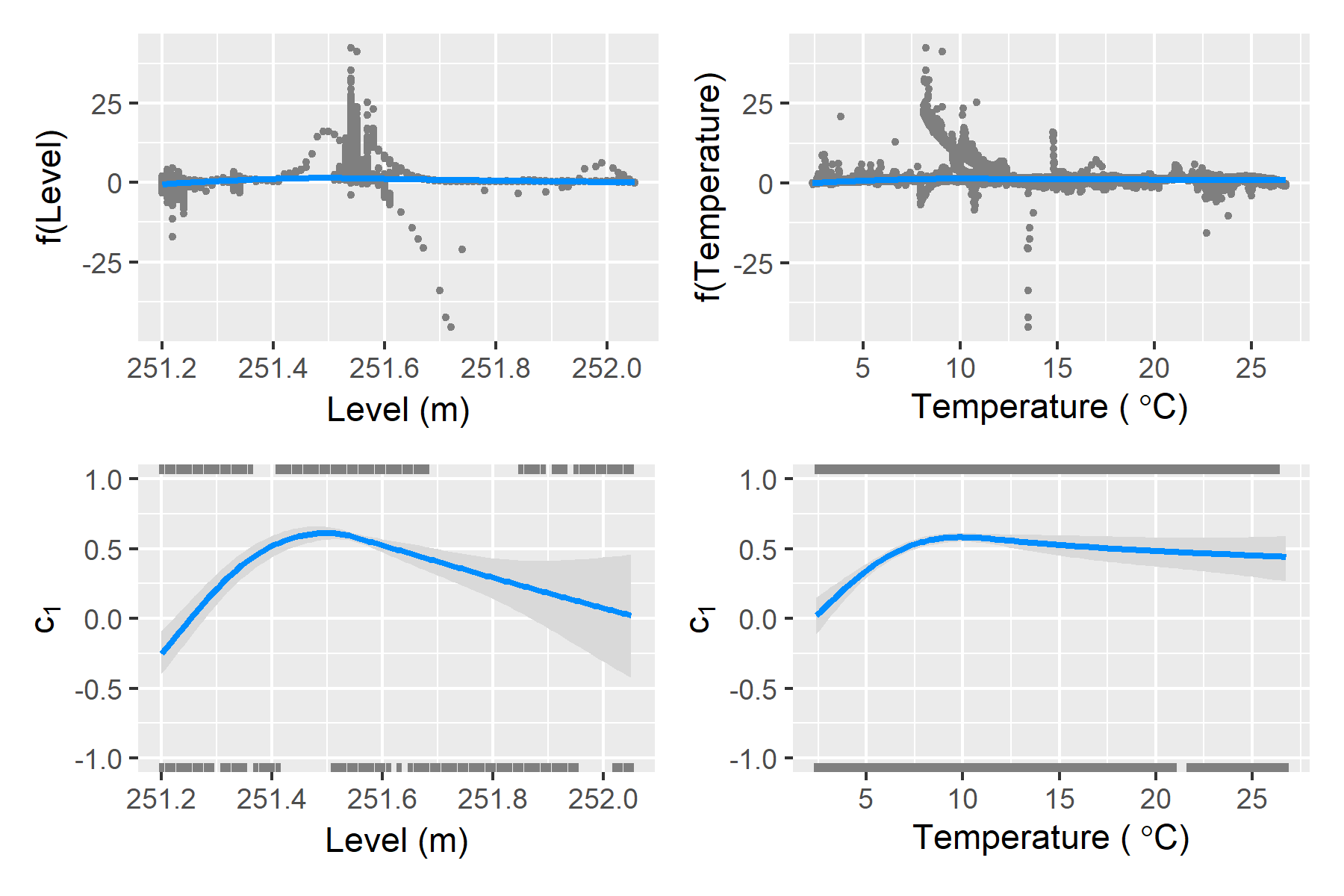} \caption{Visualizing the fitted smooth functions for conditional cross-correlation between turbidity-upstream and turbidity-downstream at lag 1 with the predictors, water level and temperature from upstream sensor. Each plot visualizes the relationship between the response and predictor while holding other predictors at their medians ($251.6m$ and $9.926^{o}C$ for water level and temperature in upstream respectively). The top panel shows the smooth terms in the predictor scale whereas the bottom panel is in the response scale.}\label{fig:visualize-ccf}
\end{figure}

At lag 1, when temperature is greater than \(10^\circ\)C it is slightly negatively affecting the cross-correlation between turbidity upstream and downstream while controlled for water level (see \autoref{fig:visualize-ccf}). Plots that visualize relationships at other lags can be obtained similarly.

\hypertarget{lag-time-prediction}{%
\subsection{Lag time prediction}\label{lag-time-prediction}}

As we described in Equation \eqref{eq:dt}, lag time is defined as the lag that gives the maximum cross-correlation conditional on the upstream variables observed at time \(t\).
This will allow \(d_t\) to vary according to the upstream river behavior.

\begin{figure}
\includegraphics[width=1\linewidth]{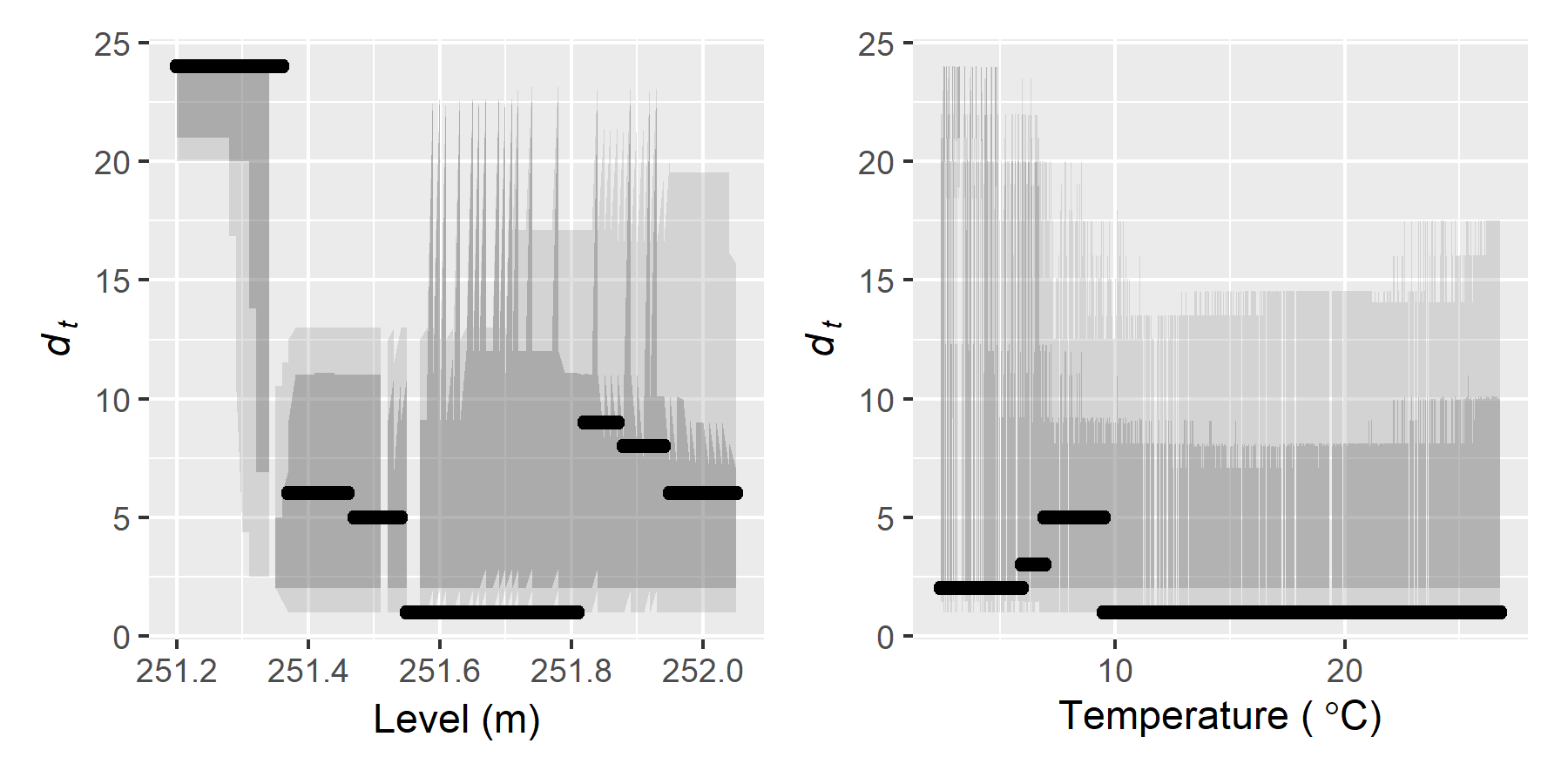} \caption{Visualizing $d_t$ with $80\%$ and $95\%$ bootstrap confidence intervals. Each panel visualizes $d_t$ vs each upstream covariate while holding the remaining upstream covariates at their medians ($251.6m$ and $9.926^{o}C$ for water level and temperature, respectively).}\label{fig:visualise-dt-withbs}
\end{figure}

The \(80\%\) and \(95\%\) confidence intervals for the relationship between estimated \(d_t\) and each upstream covariate \(z_{i,t}\) used in our conditional cross-correlation models (see \autoref{fig:visualise-dt-withbs}) are computed using the Sieve bootstrap approach \autocite{Buhlmann1997} and the algorithm is described in \autoref{sec:SieveBSalgo}.

\begin{figure}
\includegraphics[width=1\linewidth]{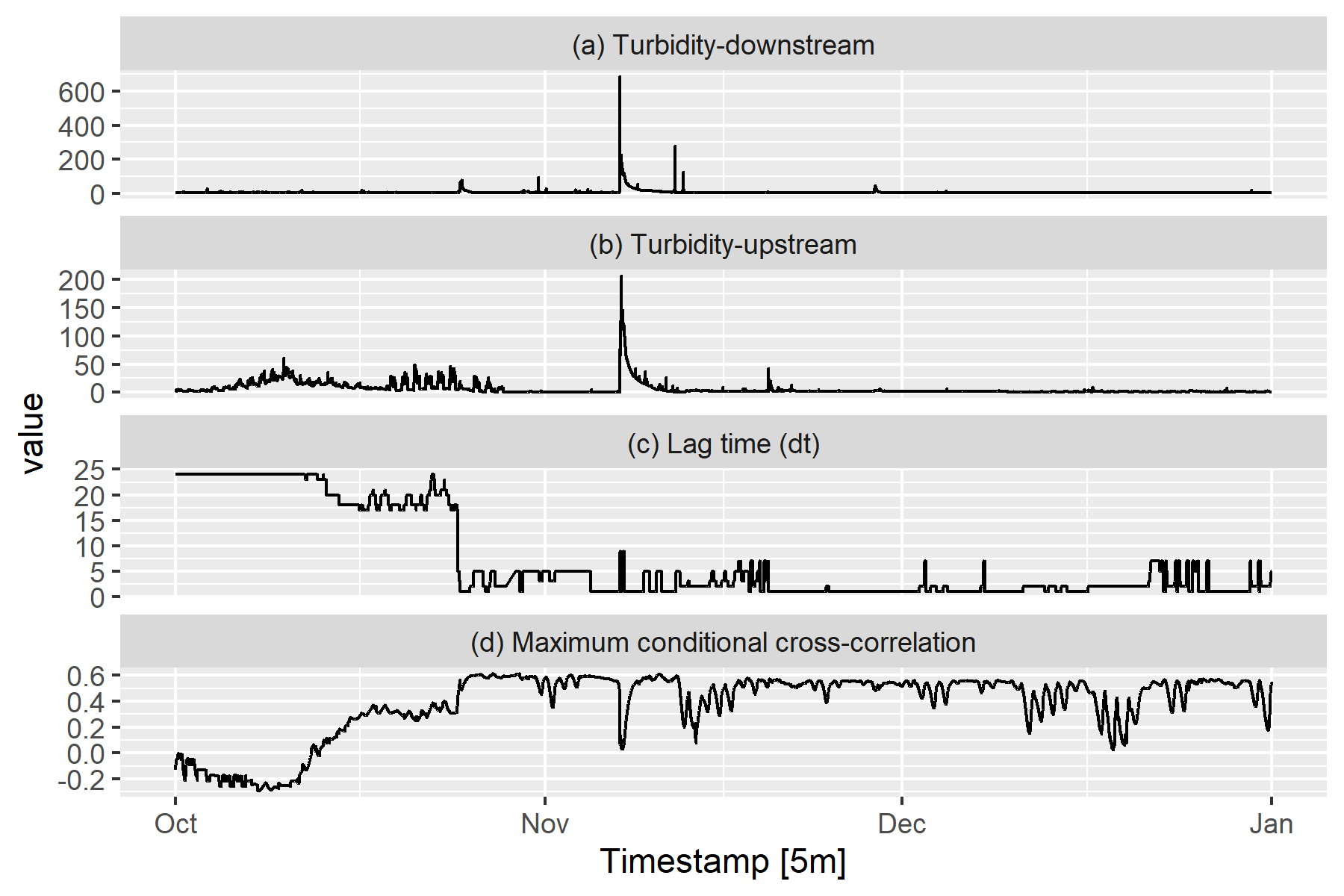} \caption{Time series plots of water-quality variables and lag time between upstream and downstream sensors for the period 01-Oct-2019 to 31-Dec-2019. (a) Time series plot of turbidity-downstream. (b) time series plot of turbidity-upstream. (c) Lag time between upstream and downstream sensors with the predictors, water level and temperature from upstream sensor (d) Maximum conditional cross-correlation between upstream and downstream sensors with the predictors, water level and temperature from upstream sensor}\label{fig:visualise-dt-with-data}
\end{figure}

To visualize the relationship between \(d_t\) and each covariate, we replace the remaining covariates with their medians in the original data, and then estimate \(d_t\) from the fitted model using this modified data. \autoref{fig:visualise-dt-withbs} displays the relationship between \(d_t\) and each covariate. It is clear from \autoref{fig:visualise-dt-withbs} that upstream water level has a negative effect on the lag time. That is, when water level increases, the lag time decreases. Increasing water level implies high fresh water inputs and more flow, water moving downstream in less time, hence the lag time will be decreasing. When the water level is between \(251.6\) and \(251.8\) \(m\), the lag time is very low. In fact, there was only one incident in November that showed a water level within this range, which occurred during a freshwater inflow event. However, when the water level is more than \(251.8\) \(m\), the lag time has increased deviating from its previous pattern. It is unclear what exactly happens at that instance, however, the original data indicates that the water level was higher than \(251.8\) \(m\) only during a single event that happened in November-\(2019\) (see \autoref{fig:timeplots-alldata} in \autoref{app:impute-turb}).
On the other hand, when the upstream temperature is below \(10^{o}C\), it has a positive effect on the lag time - that is, when the water temperature increases, the lag time also increases. This pattern is consistent with river behavior, as water temperature can increase during dry seasons when there is less inflow to the system, particularly if dry seasons occur in the warmer months. Low inflow causes the water to move downstream more slowly, resulting in an increase in the lag time. However, for temperatures greater than \(10^{o}C\), which mostly occur during early October and freshwater inflow events, the lag time remains consistently low. \autoref{fig:visualise-dt-with-data} maximum conditional cross-correlation and the lag time between turbidity-upstream and turbidity-downstream with the predictors, water level and temperature from upstream sensors.

\hypertarget{evaluation}{%
\subsection{Evaluation}\label{evaluation}}

\begin{figure}
\includegraphics[width=1\linewidth]{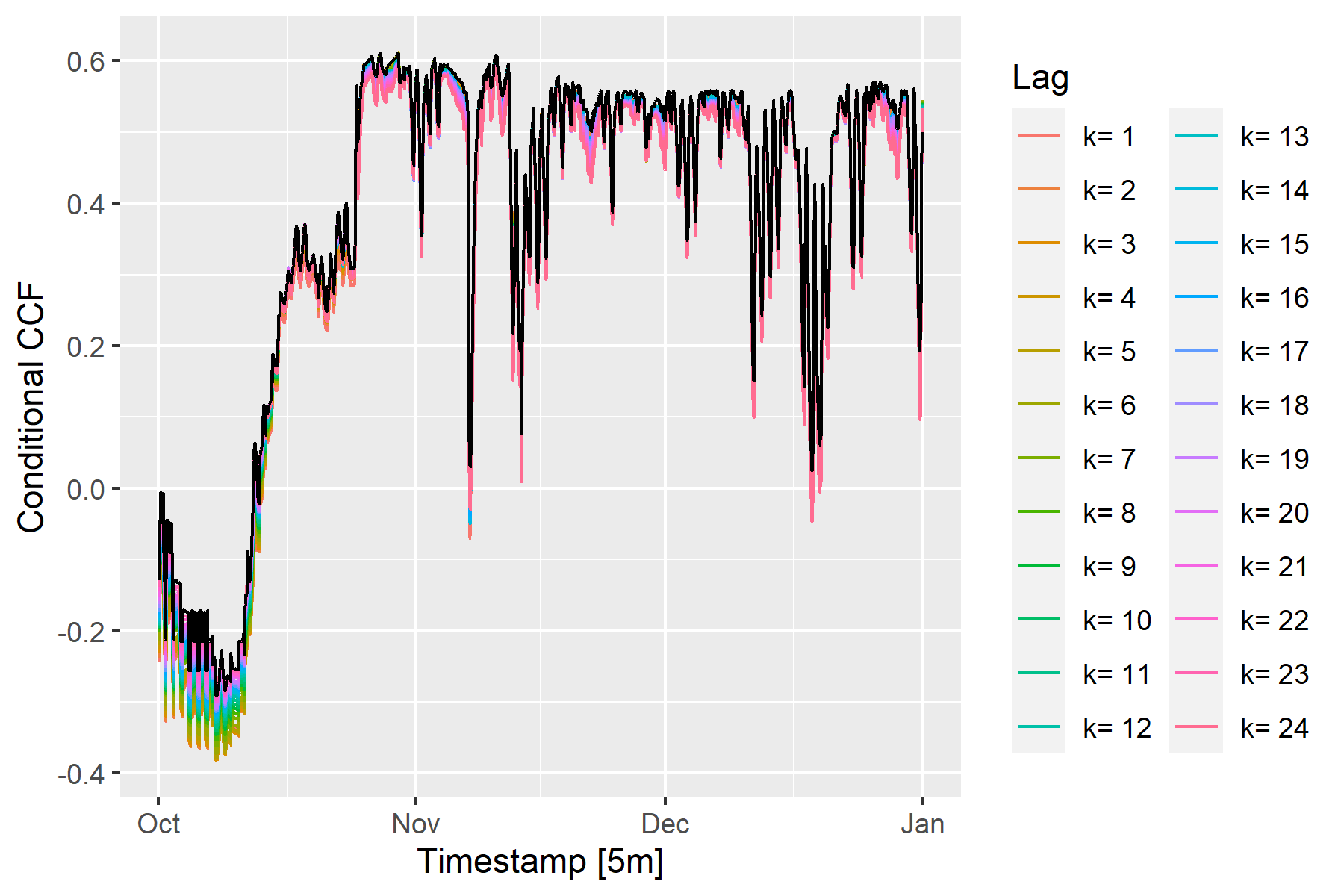} \caption{Time plot of the conditional ccf estimated at different lags,  i.e., $E[x^*_ty^*_{t+k}\mid z_t]$ for $k=1,\dots,24$. The black line represent the conditional ccf at lag $d_t$,i.e.,$E[x^*_ty^*_{t+dt}\mid z_t]$. Approximately, $96\%$ of the time $E[x^*_ty^*_{t+dt}\mid z_t] > E[x^*_ty^*_{t+k}\mid z_t]$.}\label{fig:tplot-ccf-k}
\end{figure}

We can use the estimated \(d_t\) to compute the lead variable, i.e., \(y_{t+d_t}\) (lag variable, i.e., \(x_{t-d_t}\)) from the downstream (upstream) sensor. \(y_{t+d_t}\) is expected to have the maximum conditional cross-correlation with \(x_t\) compared to any \(y_{t+k}\) for \(k=1,\dots,24\). That is, ideally we expect \(\text{E}[y^*_{t+d_t}x^*_t\mid\bm{z}_t] > \text{E}[y^*_{t+k}x^*_t\mid\bm{z}_t]\) for all lags \(k\) and time \(t\), where \(x^*_t\) and \(y^*_{t+d_t}\) are the conditionally normalized series of \(x_t\) and \(y_{t+d_t}\) with respect to \(\bm{z}_t\). To evaluate this, we first fit a GAM to \(y^*_{t+d_t}x^*_t\) using \(\bm{z}_t\) as the predictors and follow the \autoref{sec:cross-correlation} to compute \(\text{E}[y^*_{t+d_t}x^*_t\mid\bm{z}_t]\). These conditional cross-correlations are then compared with \(\text{E}[y^*_{t+k}x^*_t\mid\bm{z}_t]\) for all \(k\) and \(t\), which were obtained using Equation \eqref{eq:ccf_k}. The resultant conditional cross-correlations are shown in \autoref{fig:tplot-ccf-k}. We can see that the \(\text{E}[y^*_{t+d_t}x^*_t\mid\bm{z}_t]\) is greater than \(\text{E}[y^*_{t+k}x^*_t\mid\bm{z}_t]\) for majority of the time.

\hypertarget{sec:discussion}{%
\section{Discussion}\label{sec:discussion}}

In this study we introduce a novel approach to normalize univariate time series conditional on a set of covariates. The proposed approach uses generalized additive models to estimate the conditional mean and variance of the time series, given a set of covariates. The conditional mean is estimated via an additive model fitted to the time series with respect to the covariates. The residuals from this model are then used to estimate the conditional variance, by fitting a separate generalized additive model to the squared residuals from the conditional mean model using the same set of covariates. We assume a gamma family with a log link in the latter model. The estimated conditional means and variances are then used to normalize the time series.

Normalizing a given time series in this manner will reduce some of the variation induced through the covariates. Thus it will help to effectively model the autocorrelation of the series via appropriate time series models. Using an empirical application, we have shown that these normalized time series can be used to impute missing values and make predictions in stream temperature.

The conditionally normalized time series can also be used to compute conditional autocorrelation and conditional cross-correlation functions at different lags. To compute conditional ACF at lag \(k\), we have proposed to fit an additive model to the cross product of the normalized time series and its lagged series at \(k\), using the same set of covariates used in the normalization. Similarly, the conditional CCF at \(k\) can be estimated via an additive model fitted to the cross product of two conditionally normalized time series at \(k\) lags apart.

We have further shown that the conditional cross-correlations can be used to estimate the water travel time between two locations in a river. This lag time between two river locations varies in response to the upstream river behavior. Thus we proposed to estimate this lag time conditional on the upstream river behavior as observed by the water-quality variables measured at the upstream location. We first computed the cross-correlation between the same water-quality variable measured at both upstream and downstream locations at different lags, conditional on a set of water-quality variables measured at the upstream location. Then the lag time is computed as the lag that gives the maximum conditional cross-correlation. The significance of the maximum conditional cross-correlation was evaluated in a probabilistic way by computing standard errors of the predictions in the link space and then computing t statistics. We used this approach to estimate the water travel time between two locations in Pringle Creek, one of the NEON aquatic sites located in Texas, USA. The results show that the estimated time lag captures the highest correlation between the two water-quality variables measured at upstream and downstream locations. Lag time estimation using the conditional behavior of the river and the correlation between variables is useful in developing statistical methods for predicting other water-quality variables of interest such as sediment and nutrient concentrations in river networks \autocite{leigh2019}. Such data-driven approaches are also useful to complement or replace expensive, and time-consuming field-based methods such as salt tracer experiments.

Further research could extend these approaches, for example, considering vector autoregressions for multivariate time series problems. Similarly, models can be extended to account for spatial dependence between sites.

\hypertarget{reproducibility}{%
\section*{Reproducibility}\label{reproducibility}}
\addcontentsline{toc}{section}{Reproducibility}

All code to reproduce the results in this paper are available at \url{https://github.com/PuwasalaG/Conditional_normalisation_in_TSA}. All analysis was conducted using R \autocite{Rsoftware} and Stan \autocite{Stan}. The methods discussed are available in the conduits package for R \autocite{Rconduits}.

\hypertarget{acknowledgments}{%
\section*{Acknowledgments}\label{acknowledgments}}
\addcontentsline{toc}{section}{Acknowledgments}

This project is funded by the Australian Research Council (ARC) Linkage project (grant number: LP180101151) ``Revolutionising high resolution water-quality monitoring in the information age''. The authors acknowledge the staff members from Aquatic Instruments Science team and the National Ecological Observatory Network (NEON), especially Guy Litt, Bobby Hensley and Gary Henson, for their valuable explanations on the background of the Pringle Creek site and the relationships between water-quality variables. Further, we convey our gratitude to Erin Peterson and Claire Kermorvant for valuable discussions on the project and water-quality characteristics.

\clearpage

\appendix

\hypertarget{app:appendixA}{%
\section{Other results from the stream temperature application}\label{app:appendixA}}

\begin{figure}
\includegraphics[width=0.75\linewidth]{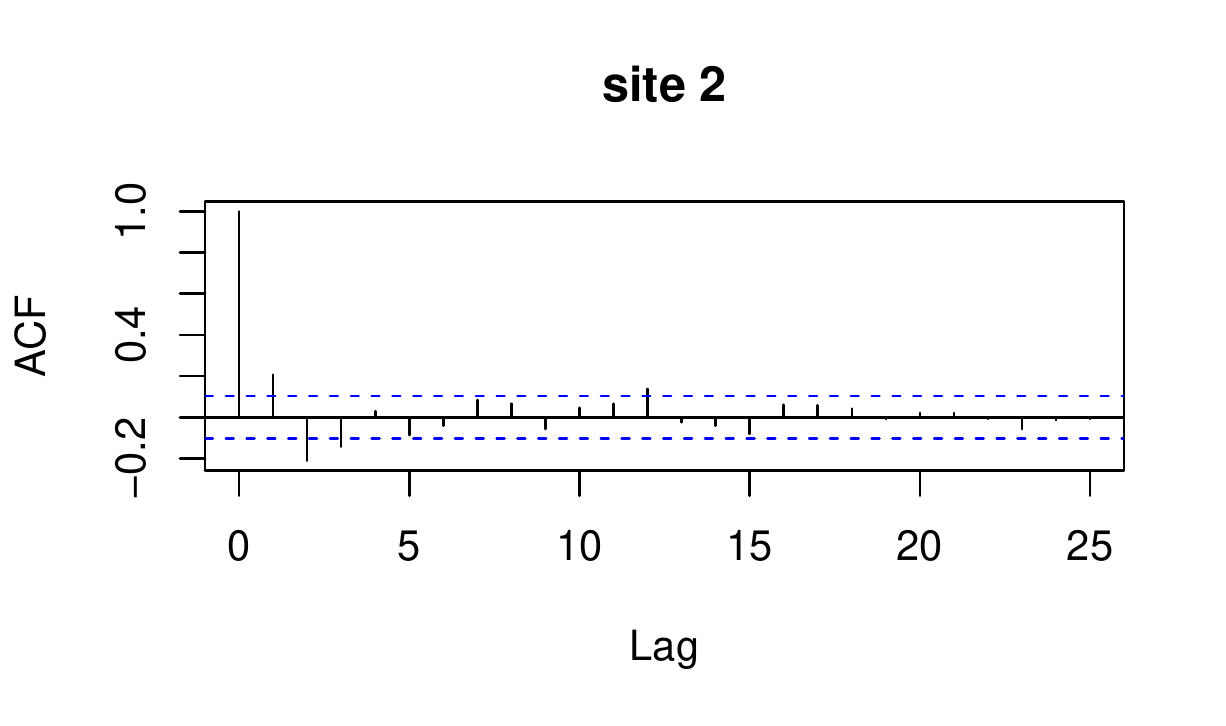} \caption{Autocorrelation plot of the standardized $y_{s = 2, t}^*$.}\label{fig:acf}
\end{figure}

\hypertarget{app:appendixB}{%
\section{Data cleaning and preliminary analysis}\label{app:appendixB}}

\hypertarget{app:pringle_creek}{%
\subsection{Study area}\label{app:pringle_creek}}

Pringle Creek drains a catchment of 48.9~km\(^2\) and experiences high flows in spring when rainfall is heaviest and low flows during the typically dry summers. Rainfall can occur in winters, typically December to January, but not snow or ice fall. The average annual temperature is about \(17.5^\circ\)C while the average annual precipitation is about 898~mm.

\begin{figure}
\includegraphics[width=1\linewidth]{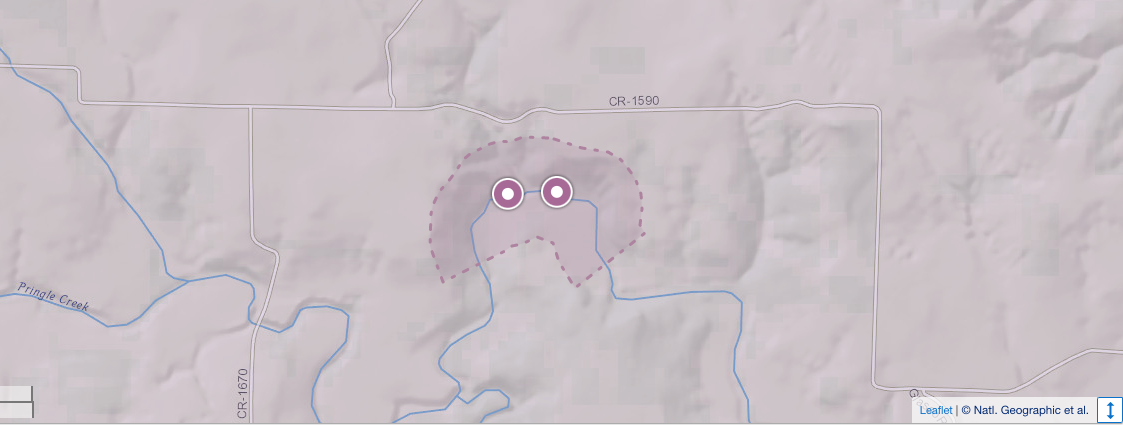} \caption{Pringle Creek sensor locations. The creek is shown in the blue line and the two pink circles denote the upstream and downstream sensor locations. Image courtesy National Ecological Observatory Network.}\label{fig:pringle-creek-map}
\end{figure}

\hypertarget{app:data_cleaning}{%
\subsection{Data cleaning}\label{app:data_cleaning}}

\begin{figure}
\includegraphics[width=1\linewidth]{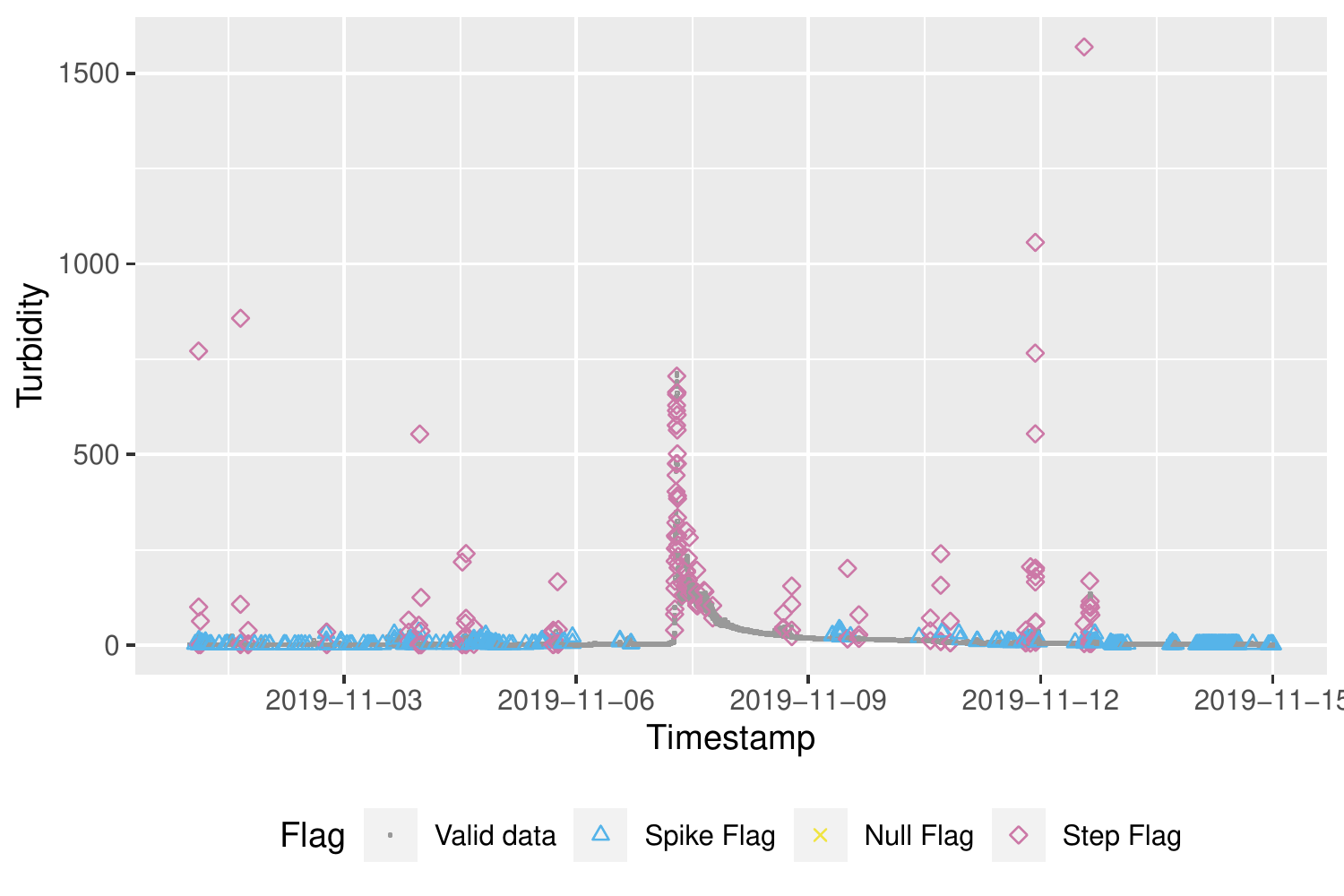} \caption{Turbidity at downstream sensor colored by the quality flags. Note that only turbidity series had technical anomalies. Other series did not show any technical anomalies during the study period we chose.}\label{fig:turbidity-downqaflags}
\end{figure}

\begin{figure}
\includegraphics[width=1\linewidth]{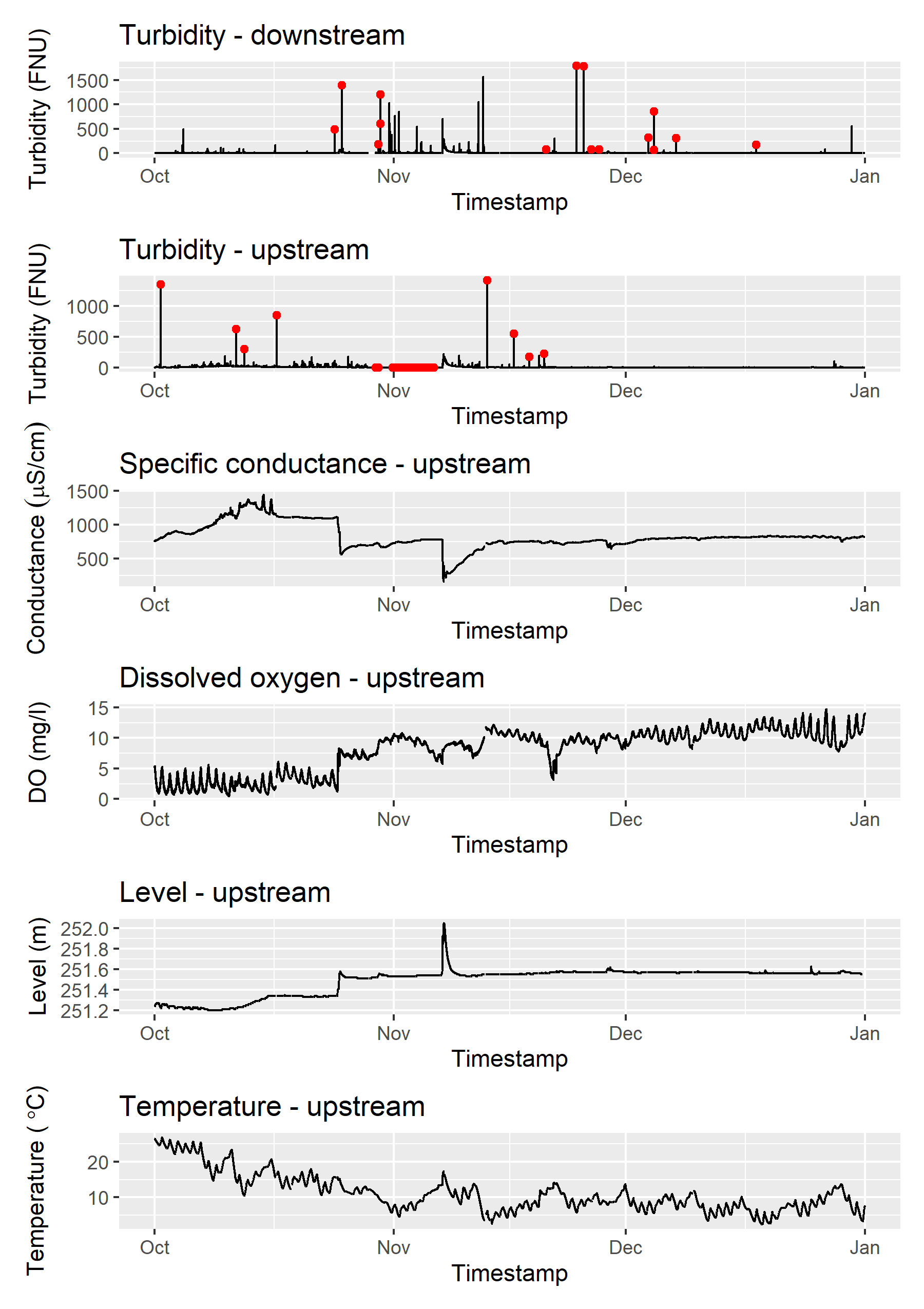} \caption{Time plots of the variables used in the study for the period spanning from 01-Oct-2019 to 31-Dec-2019. The anomalous points we identify were colored in red.}\label{fig:timeplots-flagged-anomalies}
\end{figure}

The NEON data we use in Application 2 (\autoref{sec:app2}) undergo an automated water-quality assuring process as a part of the rigorous NEON protocols, and are given a quality flag for each observation. The most commonly used quality flags include range flags, spike flags and step flags (see \textcite{NEON_QF} for a detailed explanation of these quality flags). Range flags are obvious technical anomalies as they identify the out of range observations of each water-quality variable. Therefore we use these range flags and claim these points as anomalies. However, other quality flags do not necessarily imply technical anomalies. For example, \autoref{fig:turbidity-downqaflags} shows a section of data for turbidity at the downstream sensor at Pringle Creek, colored by the quality flags given by the automated process. In the event of freshwater inflows, turbidity tends to increase and then gradually decrease; the natural behavior of turbidity in many river systems. However, these types of points (i.e., sudden increases in turbidity) are flagged by the automated process, even though they are unlikely to be technical anomalies. Therefore we did not directly use any quality flags other than range flags in this study, and claim the points shown in \autoref{fig:timeplots-flagged-anomalies} as technical anomalies.

\begin{figure}
\includegraphics[width=1\linewidth]{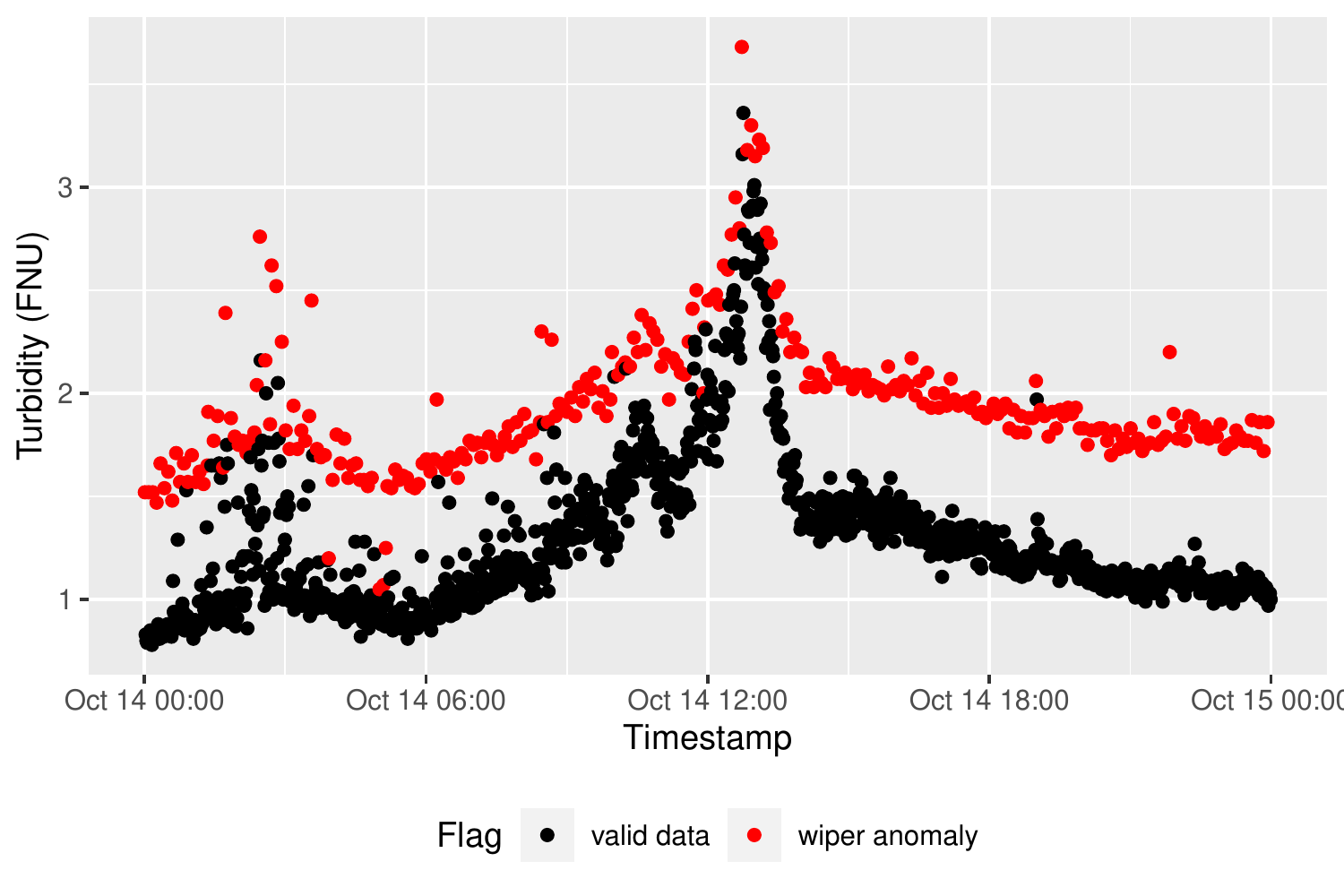} \caption{Wiper anomalies in turbidity downstream sensor. This plot only shows data for 14-Oct-2019 to distinguish the scale difference between wiper anomalies and typical data.}\label{fig:wiper-anomaly}
\end{figure}

Apart from these anomalous points, we also noticed that every fifth observation in turbidity in both sites are anomalous as a result of the wiper on the optical turbidity sensor operating (i.e.~wiping any biofouling off the sensor probe) every five minutes (\autoref{fig:wiper-anomaly}). Therefore we also discarded these anomalous points in turbidity series from both sites.

\hypertarget{app:prel_plots}{%
\subsection{Preliminary analysis}\label{app:prel_plots}}

\begin{figure}
\includegraphics[width=1\linewidth]{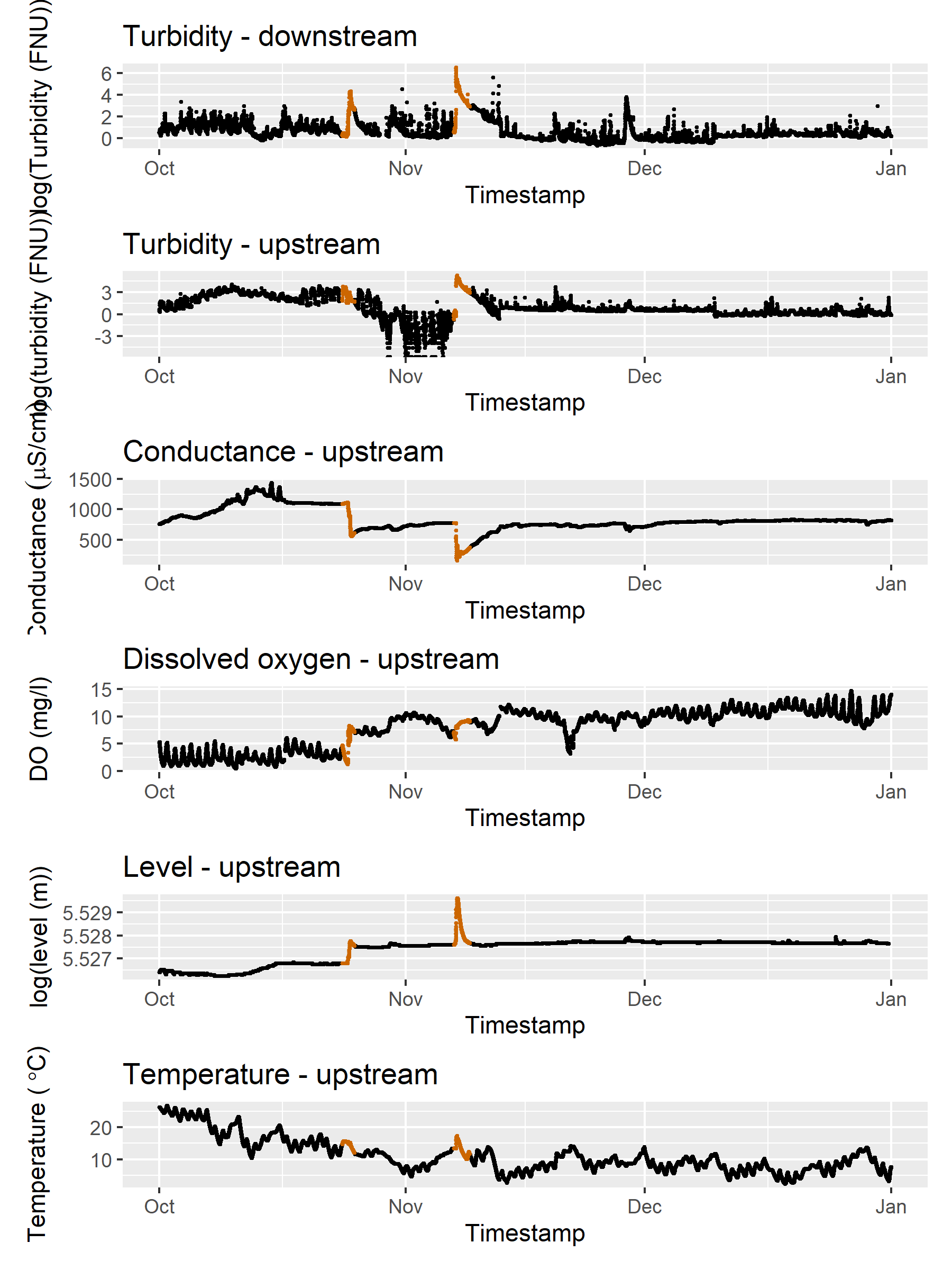} \caption{Time plots of water-quality variables for the period 01-Oct-2019 to 31-Dec-2019. All variables are aggregated into 5-minute frequencies, post anomaly removal (See \autoref{app:data_cleaning}). Turbidity and water level are shown on the log scale. Observations highlighted in orange show examples of patterns in water quality when there are fresh water inflows (water level rises).}\label{fig:timeplots-alldata-scaled}
\end{figure}

\begin{figure}
\includegraphics[width=1\linewidth]{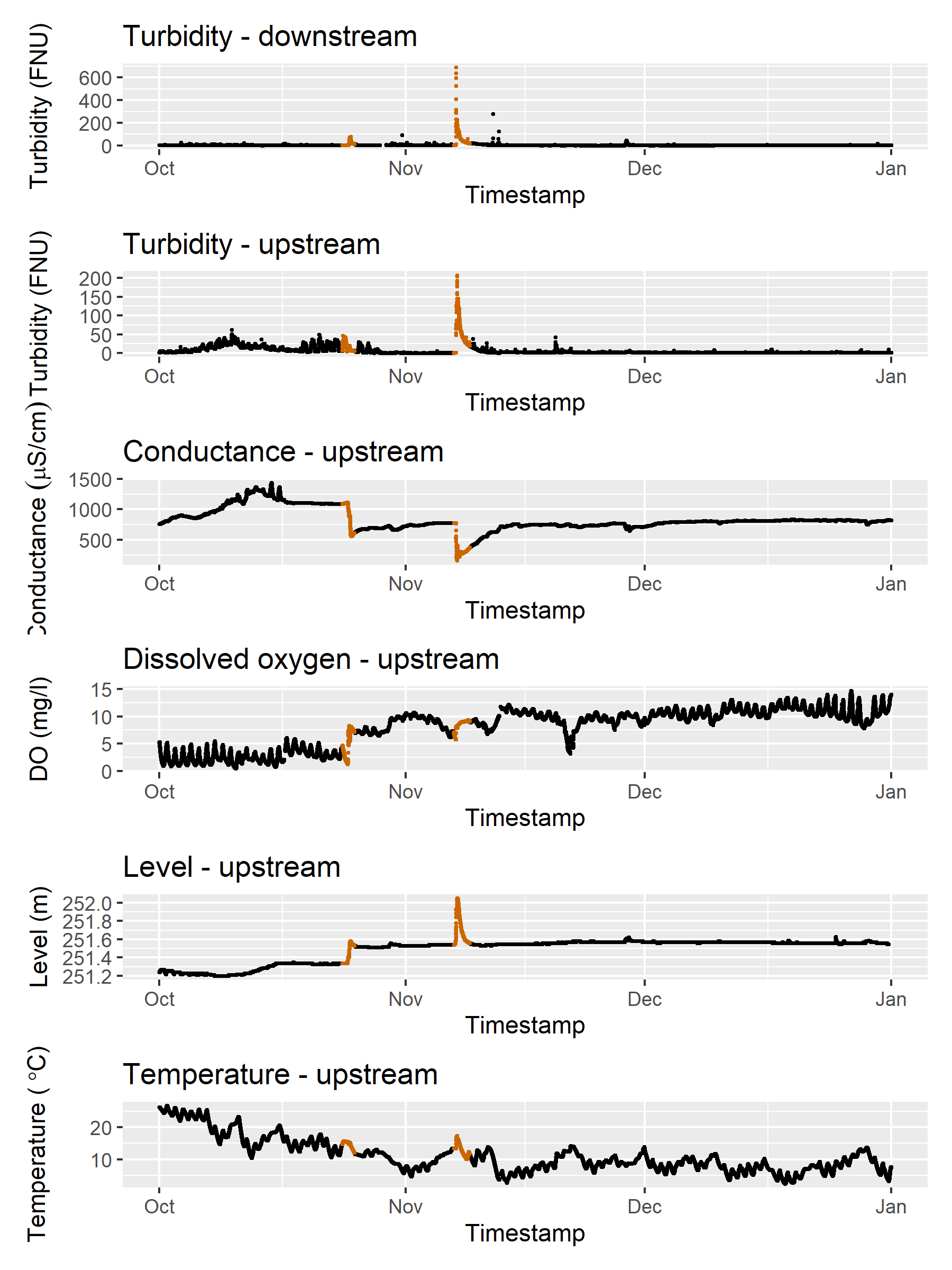} \caption{Time plots of water-quality variables for the period 01-Oct-2019 to 31-Dec-2019. All variables are aggregated into 5-minute frequencies, post anomaly removal (See Appendix \autoref{app:data_cleaning}). All water-quality variables are plotted in their original scale. A couple of instances are highlighted in orange to show the patterns in water-quality variables when there is fresh water inflows.}\label{fig:timeplots-alldata}
\end{figure}

Prior to the analysis, we removed all the anomalous points (see Appendix \ref{app:data_cleaning}), then aggregated the observations for all water-quality variables into 5-minute frequency data (\autoref{fig:timeplots-alldata-scaled}). The resultant time series reflect the typical behavior of water quality in many rivers. \autoref{fig:timeplots-alldata} is similar to \autoref{fig:timeplots-alldata-scaled} except that the turbidity and level are plotted in log scale in \autoref{fig:timeplots-alldata-scaled}. Changed scales in some variables has allowed us to see the internal variations in these individual variables.

\begin{figure}
\includegraphics[width=1\linewidth]{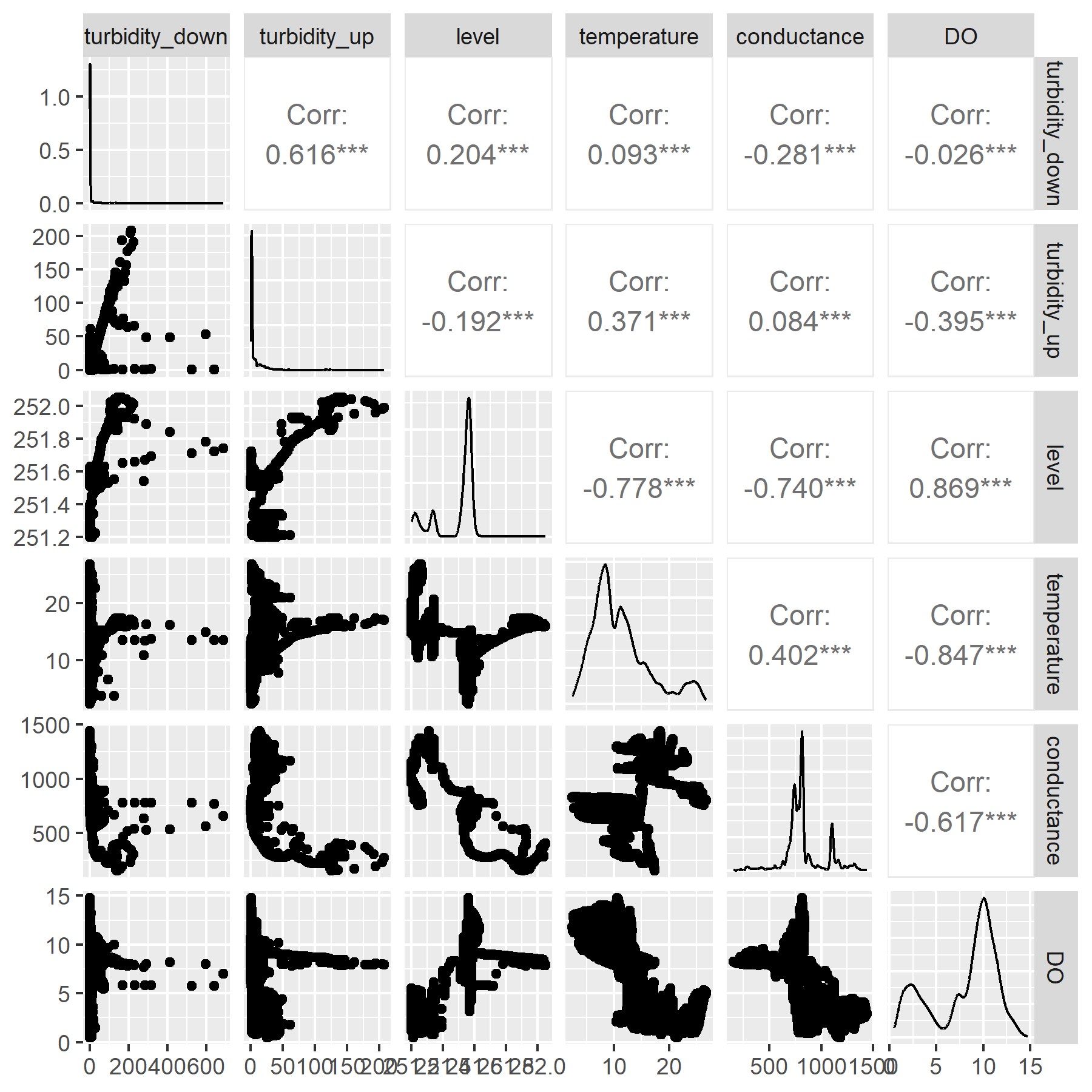} \caption{Pairwise scatter plots between the turbidity and other covariates from the upstream sensor. Upper triangular matrices shows the Pearson's correlation coefficient for each pair.}\label{fig:ggpairs-plot}
\end{figure}

Turbidity tends to increase when freshwater flows into the river (when water level rises) as this will increase the suspended particles in water. In contrast, conductance tends to decrease with fresh water inflows as the water becomes diluted \autocite{leigh2019} (see \autoref{fig:timeplots-alldata-scaled}). This behavior also explains why the relationship between water level at the upstream location and turbidity at the upstream site is much stronger than that between the upstream and downstream locations (see \autoref{fig:ggpairs-plot}).

We then chose the set of covariates from the upstream sensor to model the cross-correlation between upstream and downstream turbidity by visually analysing the relationship between the variables. From \autoref{fig:ggpairs-plot}, we can see that the upstream water level, temperature and conductance show non-linear relationships with both turbidity series. In contrast, dissolved-oxygen does not show much relationship with turbidity. We also see that water level and conductance have a strong non-linear relationship. Hence, choosing both water level and conductance as covariates could lead to multicollinearity problems. Given these observations, we chose water level and temperature as the covariates to compute conditional cross-correlations between upstream and downstream turbidity series.

\begin{figure}
\includegraphics[width=1\linewidth]{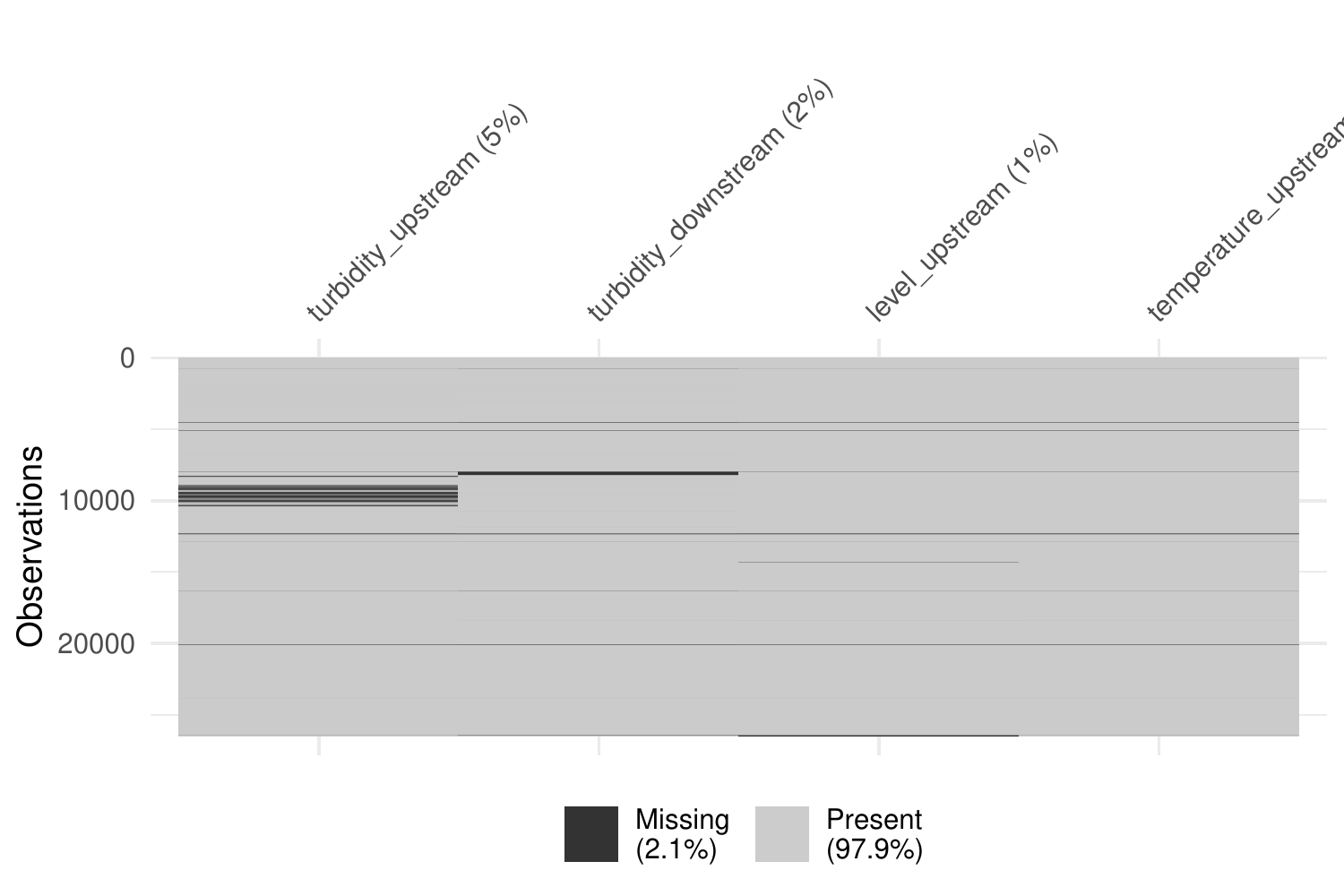} \caption{Visualization of the missing values in the variables.}\label{fig:vis-miss}
\end{figure}

\begin{figure}
\includegraphics[width=1\linewidth]{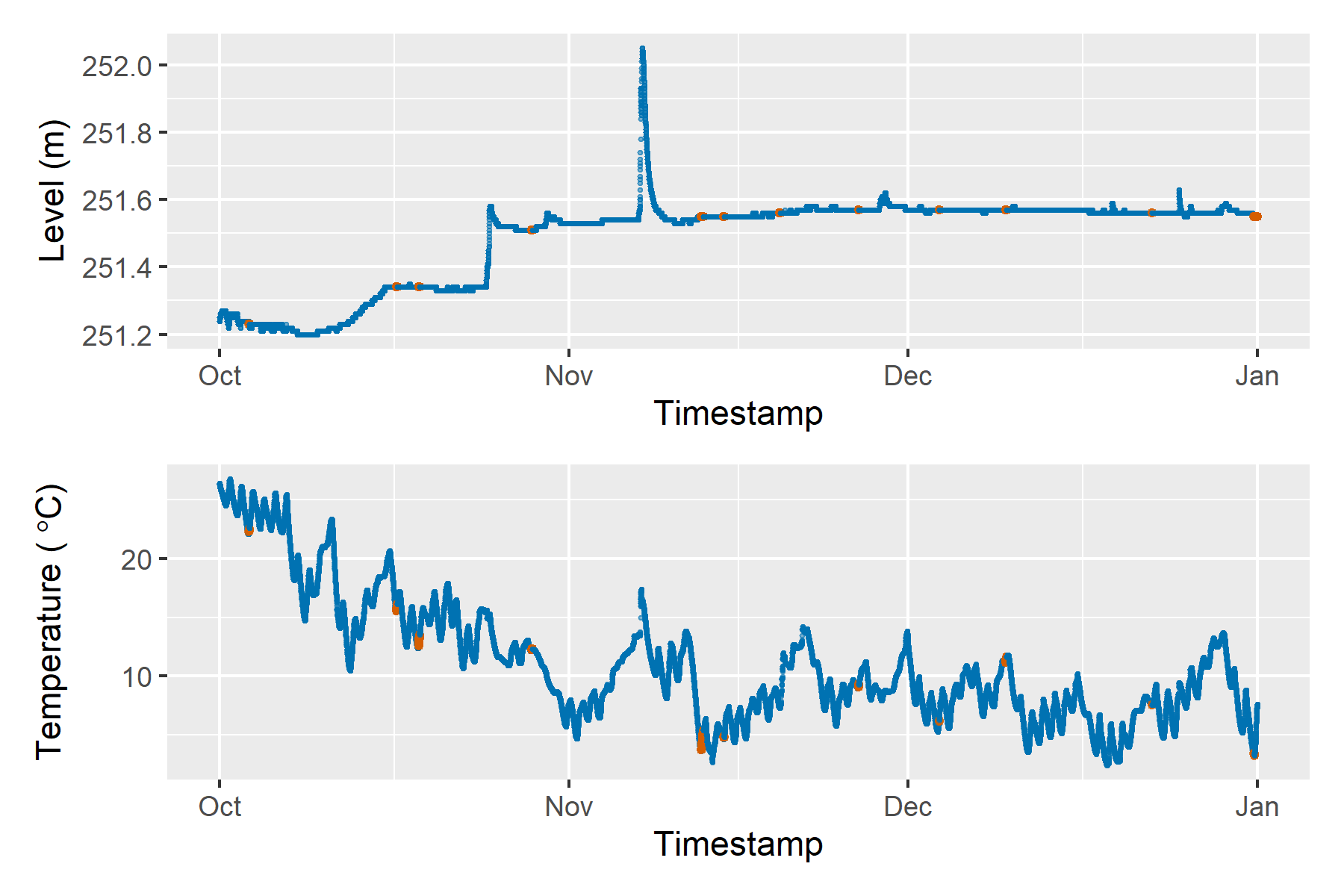} \caption{Time plot of upstream level and temperature with imputed values. Points in orange denote the imputed missing values.}\label{fig:imputed-predictors}
\end{figure}

Prior to the remaining analysis, we impute the missing values in upstream level and temperature (see \autoref{fig:vis-miss} to visualize the percentage of missing values). The missing values in water level are imputed using linear interpolation whereas the missing values in temperature are imputed using a Kalman-smoother implemented in the \texttt{imputeTS} R package \autocite{imputeTS}, based on a state space representation of the ARIMA model chosen by \texttt{auto.arima} implemented in the \texttt{forecast} R package \autocite{forecast,HK08}. \autoref{fig:imputed-predictors} plots the time series for level and temperature with the imputations.

\hypertarget{app:impute-turb}{%
\subsection{Imputing missing values of turbidity upstream via conditional normalization}\label{app:impute-turb}}

To estimate the lag time between two sensors, we compute the conditional cross-correlations, assuming that the lag time depends on
the water-quality at the upstream location. First we will normalize the two turbidity series conditional on the upstream level and temperature.

\autoref{fig:vis-miss} shows that the turbidity upstream has considerable amount of missing values which might affect the lag time estimation. Therefore, we first impute these missing values following the method explained in \autoref{sec:imputing_missing}. Let upstream turbidity is denoted by \(x_t\) and the conditional normalized series of that is given by \(x^*_t = \frac{x_{t} - \hat{m}_x(\bm{z}_t)}{\sqrt{\hat{v}_x(\bm{z}_{t})}}\) where \(\bm{z}_t\) contains the water level and temperature measured at the upstream sensor. The conditional means, \(\hat{m}_x(\bm{z}_t)\) and variances, \(\hat{v}_x(\bm{z}_t)\) are computed using Equations \eqref{eq:cond_mean} and \eqref{eq:cond_var} respectively. We use the generalized additive models implemented in the \texttt{mgcv} R package \autocite{mgcv,wood2017mgcvR}. Thin plate regression splines \autocite{wood2003TP} are used as the smooth function for each covariate. We can set the dimension \(k\) of the smoother by observing the relationship between the response and the predictor. Caution should be taken when choosing \(k\), because larger values can lead to overfitting due to the underlying autocorrelation in the time series. See \textcite{wood2017mgcvR} for a discussion on choosing \(k\) in GAMs.

\begin{figure}
\includegraphics[width=1\linewidth]{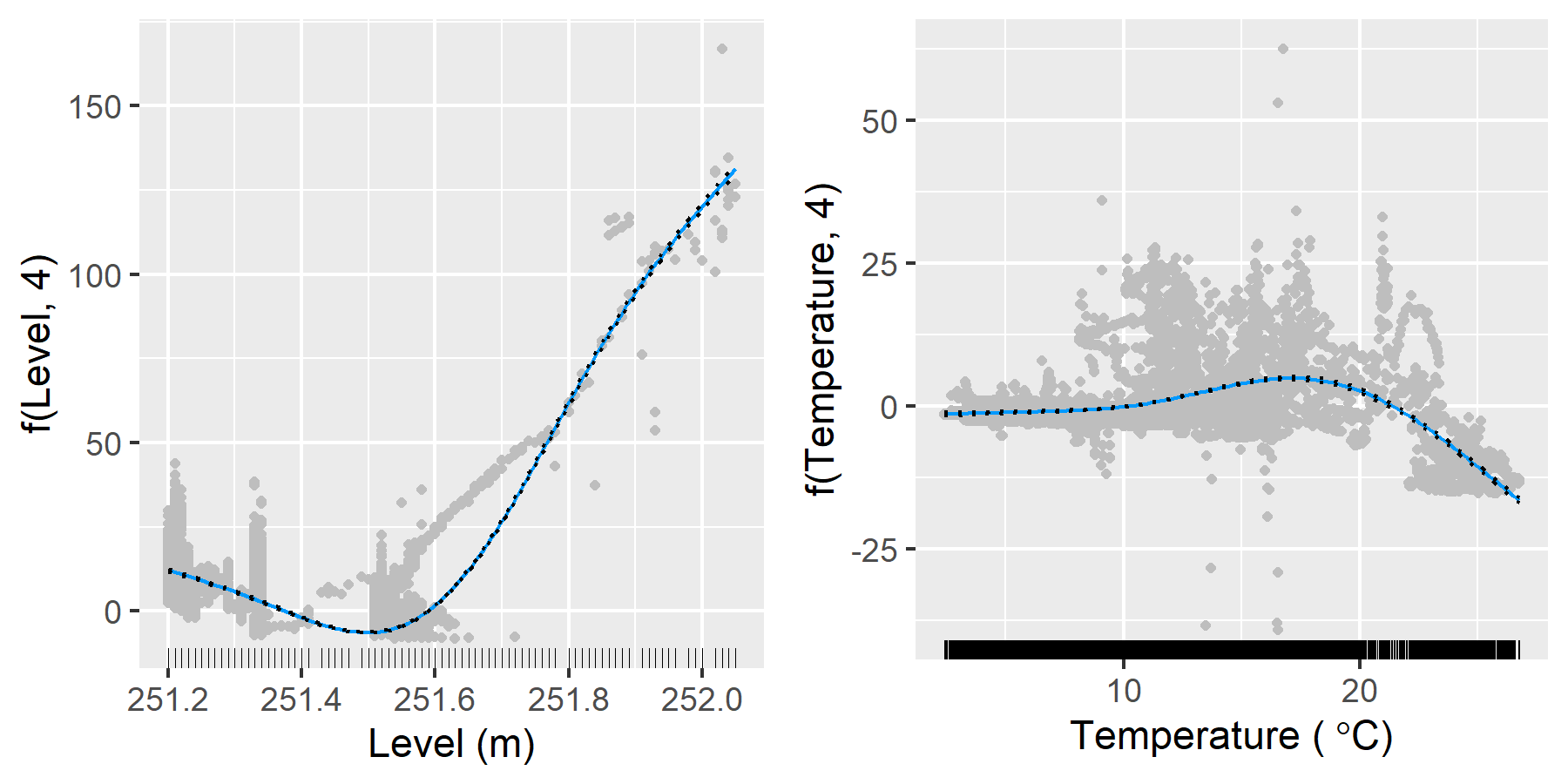} \caption{Visualizing the fitted smooth functions in the conditional mean model for turbidity upstream with the predictors, water level and temperature from upstream sensor. Each panel visualizes the relationship between the response and predictor while holding other predictors at their medians ($251.6m$ and $9.926^{o}C$ for water level and temperature, respectively). The smooth function is shown in blue and the black points are the partial residuals. The degrees of the smoothing are shown in the y-axis label for each plot.}\label{fig:cond-mean-turbidity-upstream}
\end{figure}

From \autoref{fig:cond-mean-turbidity-upstream} it can be seen that turbidity has a positive relationship with water level when adjusted for temperature and the relationship is stronger when the water level is higher. However, temperature does not seem to have much effect on turbidity when adjusted for water level.

\begin{figure}
\includegraphics[width=1\linewidth]{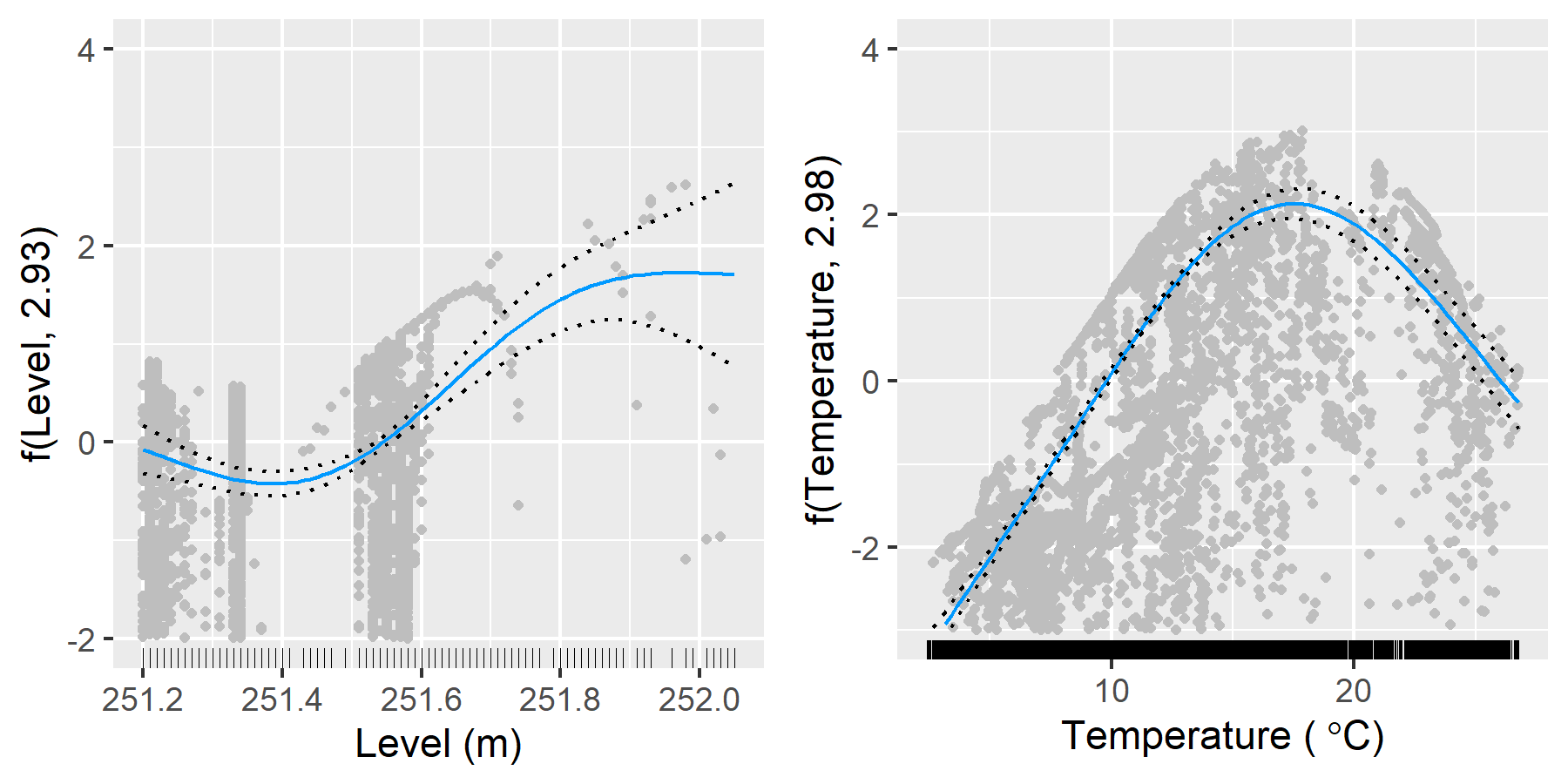} \caption{Visualizing the fitted smooth functions in the conditional variance model for turbidity upstream with the predictors, water level and temperature from upstream sensor. Each panel visualizes the relationship between the response and predictor while holding other predictors at their medians ($251.6m$ and $9.926^{o}C$ for upstream water level and temperature, respectively). The smooth function is shown in blue and the black points are the partial residuals. The degrees of the smoothing are shown in the y-axis label for each plot.}\label{fig:cond-var-turbidity-upstream}
\end{figure}

To compute the conditional variance, we model the squared residuals from the conditional mean models assuming a Gamma family with a log link as in Equation \eqref{eq:cond_var}. \autoref{fig:cond-var-turbidity-upstream} shows the relationship between the response and each predictor in the fitted conditional variance model.

\begin{figure}
\includegraphics[width=1\linewidth]{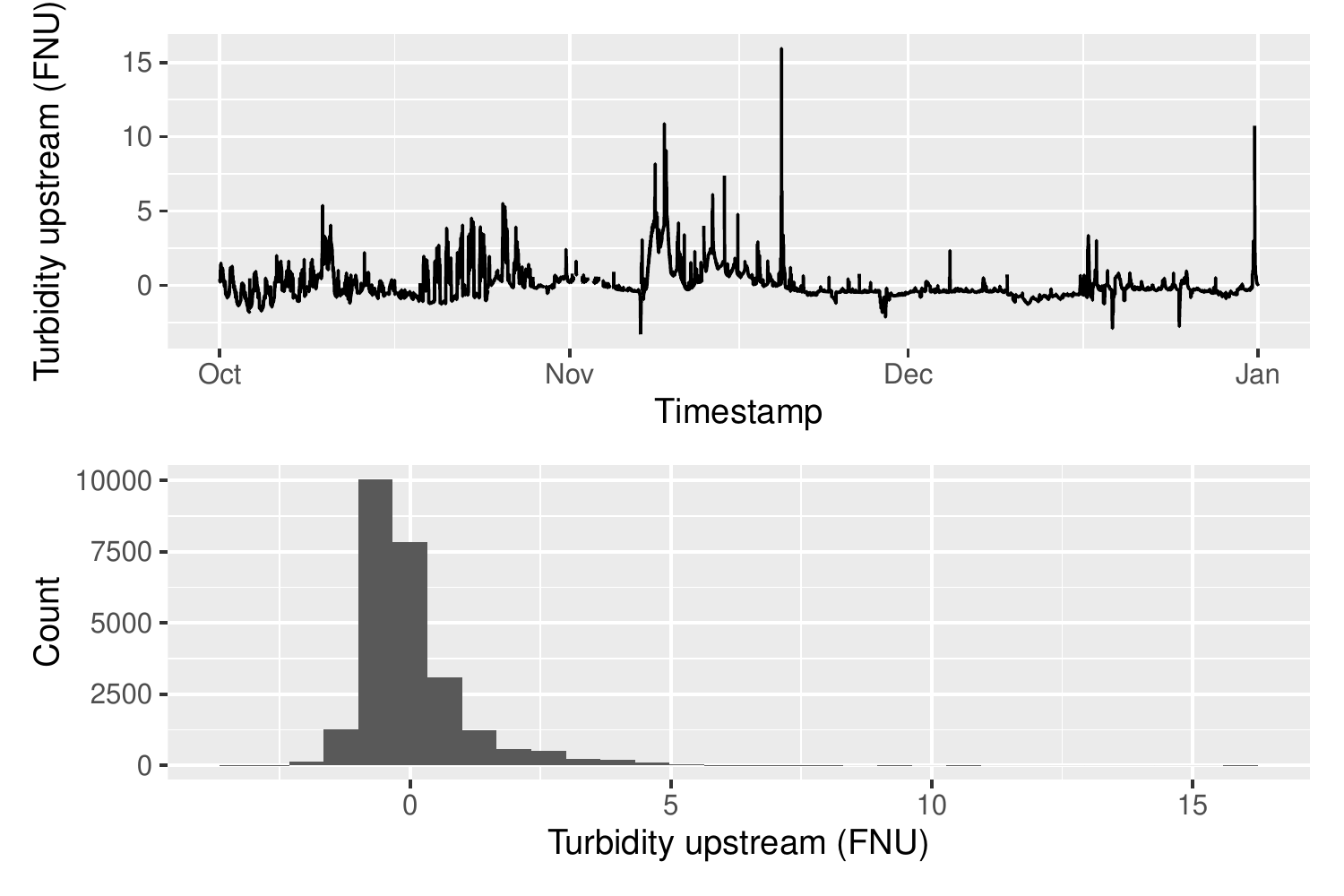} \caption{Conditionally normalized upstream and downstream turbidity.}\label{fig:normalised-turbidity-upstream-print}
\end{figure}

\begin{figure}
\includegraphics[width=1\linewidth]{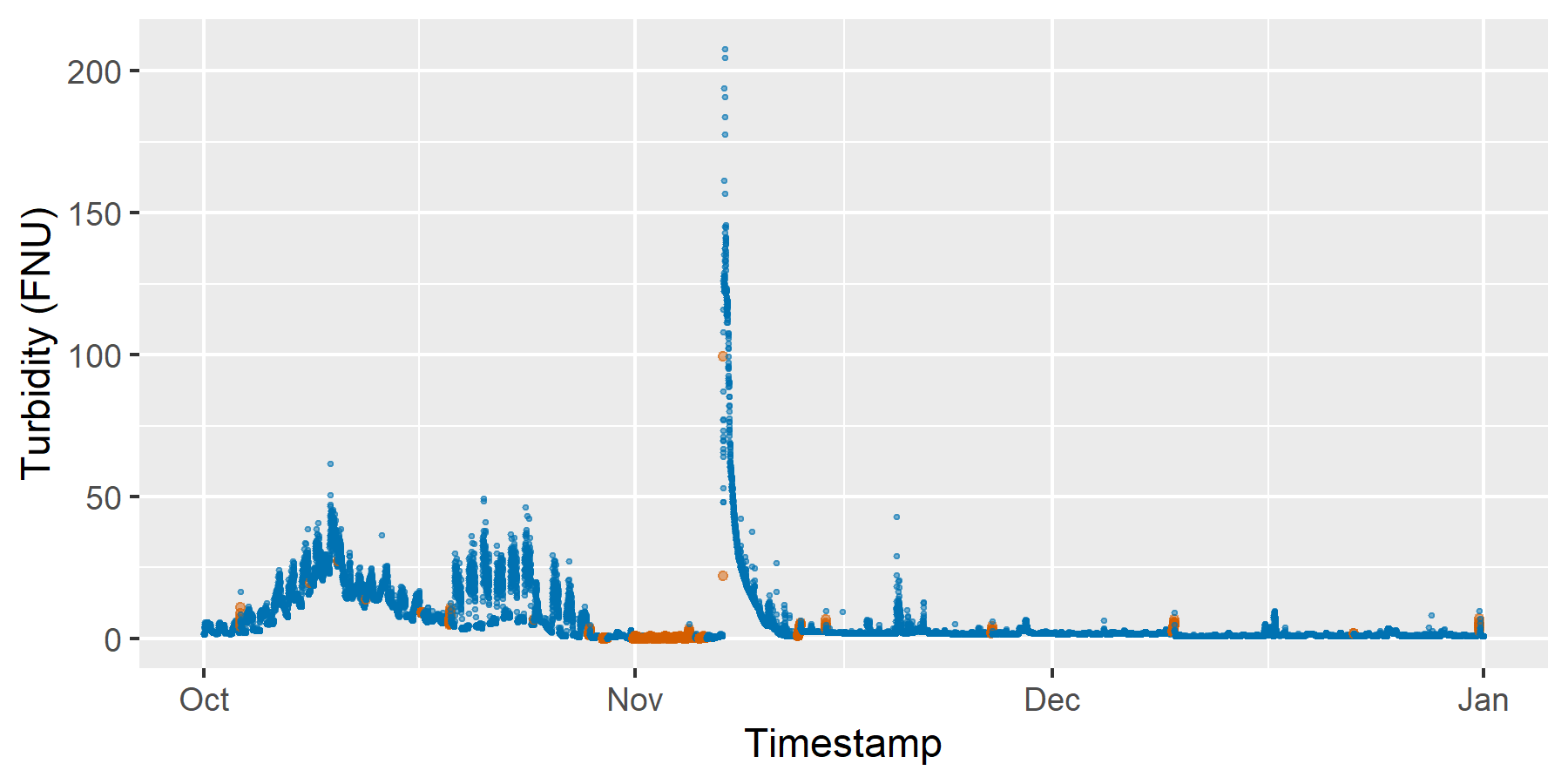} \caption{Time plot of upstream turbidity and temperature with imputed values. Points in orange denote the imputed missing values.}\label{fig:imputed-turbidity-print}
\end{figure}

The normalized series for turbidity upstream is shown in \autoref{fig:normalised-turbidity-upstream-print}. To impute the missing values of turbidity upstream, we fit an ARIMA model to \(x^*_{t}\) using the \texttt{auto.arima} function from the \texttt{forecast} R package \autocite{forecast,HK08}. Following Equation \eqref{eq:univ_interpolate} we then impute the missing values of \(x_t\) as \(\hat{x}_t^*\sqrt{\hat{v}(\bm{z}_t)} + \hat{m}(\bm{z}_t)\), where \(\hat{x}^*_t\) is computed using the Kalman smoother implemented in the \texttt{imputeTS} R package \autocite{imputeTS}. It should be noted that, this method leads to impute negative values. Since the turbidity cannot be negative, we have removed those points in the remaining analysis (see \autoref{fig:imputed-turbidity-print}).

\newpage

\printbibliography

@article{bal2014hierarchical,
  title     = {A hierarchical Bayesian model to quantify uncertainty of stream water temperature forecasts},
  author    = {Bal, Guillaume and Rivot, Etienne and Baglini{\`e}re, Jean-Luc and White, Jonathan and Pr{\'e}vost, Etienne},
  journal   = {PLoS One},
  volume    = {9},
  number    = {12},
  pages     = {e115659},
  year      = {2014},
  publisher = {Public Library of Science}
}

@article{Buhlmann1997,
  issn      = {13507265},
  url       = {http://www.jstor.org/stable/3318584},
  author    = {Peter Bühlmann},
  journal   = {Bernoulli},
  number    = {2},
  pages     = {123--148},
  publisher = {International Statistical Institute (ISI) and Bernoulli Society for Mathematical Statistics and Probability},
  title     = {Sieve Bootstrap for Time Series},
  volume    = {3},
  year      = {1997}
}

@manual{forecast,
  title  = {{forecast}: Forecasting functions for time series and linear models},
  author = {Rob Hyndman and George Athanasopoulos and Christoph Bergmeir and Gabriel Caceres and Leanne Chhay and Mitchell O'Hara-Wild and Fotios Petropoulos and Slava Razbash and Earo Wang and Farah Yasmeen},
  year   = {2022},
  note   = {R package version 8.16},
  url    = {https://pkg.robjhyndman.com/forecast/}
}

@book{fpp3,
  title     = {Forecasting: principles and practice},
  author    = {Rob J Hyndman and George Athanasopoulos},
  year      = 2021,
  publisher = {OTexts},
  url       = {OTexts.org/fpp3},
  edition   = {3rd ed},
  address   = {Melbourne, Australia}
}

@article{Green2002,
  title     = {Calculation of time of concentration for hydrologic design and analysis using geographic information system vector objects},
  author    = {Green, Jonathan I and Nelson, E James},
  journal   = {Journal of Hydroinformatics},
  volume    = {4},
  number    = {2},
  pages     = {75--81},
  year      = {2002},
  publisher = {IWA Publishing}
}

@book{hastie1990generalized,
  title     = {Generalized additive models},
  author    = {Hastie, Trevor J and Tibshirani, Robert J},
  volume    = {43},
  year      = {1990},
  publisher = {CRC press}
}

@article{HK08,
  title   = {Automatic time series forecasting: the forecast package for {R}},
  author  = {Rob J Hyndman and Yeasmin Khandakar},
  year    = 2008,
  number  = {3},
  pages   = {1--22},
  volume  = {26},
  journal = {Journal of Statistical Software}
}

@article{hrachowitz2016,
  title     = {Transit times—The link between hydrology and water quality at the catchment scale},
  author    = {Hrachowitz, Markus and Benettin, Paolo and Van Breukelen, Boris M and Fovet, Ophelie and Howden, Nicholas JK and Ruiz, Laurent and Van Der Velde, Ype and Wade, Andrew J},
  journal   = {Wiley Interdisciplinary Reviews: Water},
  volume    = {3},
  number    = {5},
  pages     = {629--657},
  year      = {2016},
  publisher = {Wiley Online Library}
}

@article{imputeTS,
  title   = {{imputeTS: Time Series Missing Value Imputation in R}},
  author  = {Steffen Moritz and Thomas Bartz-Beielstein},
  journal = {{The R Journal}},
  volume  = {9},
  number  = {1},
  pages   = {207--218},
  year    = {2017},
  doi     = {10.32614/RJ-2017-009}
}

@article{isaak2017norwest,
  title     = {The NorWeST summer stream temperature model and scenarios for the western US: A crowd-sourced database and new geospatial tools foster a user community and predict broad climate warming of rivers and streams},
  author    = {Isaak, Daniel J and Wenger, Seth J and Peterson, Erin E and Ver Hoef, Jay M and Nagel, David E and Luce, Charles H and Hostetler, Steven W and Dunham, Jason B and Roper, Brett B and Wollrab, Sherry P and others},
  journal   = {Water Resources Research},
  volume    = {53},
  number    = {11},
  pages     = {9181--9205},
  year      = {2017},
  publisher = {Wiley Online Library}
}

@article{leigh2019,
  title     = {Predicting sediment and nutrient concentrations from high-frequency water-quality data},
  author    = {Leigh, Catherine and Kandanaarachchi, Sevvandi and McGree, James M and Hyndman, Rob J and Alsibai, Omar and Mengersen, Kerrie and Peterson, Erin E},
  journal   = {PloS one},
  volume    = {14},
  number    = {8},
  pages     = {e0215503},
  year      = {2019},
  publisher = {Public Library of Science San Francisco, CA USA}
}

@article{li2008,
  title     = {Overland flow time of concentration on very flat terrains},
  author    = {Li, Ming-Han and Chibber, Paramjit},
  journal   = {Transportation Research Record},
  volume    = {2060},
  number    = {1},
  pages     = {133--140},
  year      = {2008},
  publisher = {SAGE Publications Sage CA: Los Angeles, CA}
}

@article{li2018,
  title   = {The time delay of flow and sediment in the Middle and Lower Yangtze River and its response to the Three Gorges Dam},
  author  = {Li, Yangyang and Zhu, Yingxin and Chen, Lei and Shen, Zhenyao},
  journal = {Journal of Hydrometeorology},
  volume  = {19},
  number  = {3},
  pages   = {625--638},
  year    = {2018}
}

@manual{mgcv,
  title  = {{mgcv}: Mixed GAM Computation Vehicle with Automatic Smoothness
            Estimation},
  author = {Simon Wood},
  year   = {2020},
  note   = {R package version 1.8-33},
  url    = {https://cran.r-project.org/package=mgcv}
}

@misc{NEON_data_level,
  doi       = {10.48443/QSER-8M94},
  url       = {https://data.neonscience.org/data-products/DP1.20016.001/RELEASE-2021},
  author    = {{National Ecological Observatory Network (NEON)}},
  language  = {en},
  title     = {Elevation of surface water (DP1.20016.001)},
  publisher = {National Ecological Observatory Network (NEON)},
  year      = {2021}
}

@misc{NEON_data_temp,
  doi       = {10.48443/NY19-PJ91},
  url       = {https://data.neonscience.org/data-products/DP1.20053.001/RELEASE-2021},
  author    = {{National Ecological Observatory Network (NEON)}},
  language  = {en},
  title     = {Temperature (PRT) in surface water (DP1.20053.001)},
  publisher = {National Ecological Observatory Network (NEON)},
  year      = {2021}
}

@misc{NEON_data_WQ,
  doi       = {10.48443/D8KW-5J62},
  url       = {https://data.neonscience.org/data-products/DP1.20288.001/RELEASE-2021},
  author    = {{National Ecological Observatory Network (NEON)}},
  language  = {en},
  title     = {Water quality (DP1.20288.001)},
  publisher = {National Ecological Observatory Network (NEON)},
  year      = {2021}
}

@misc{NEON_QF,
  doi       = {},
  url       = {https://data.neonscience.org/api/v0/documents/NEON.DOC.004931vB},
  author    = {Cawley, Kaelin M},
  title     = {NEON Algorithm Theoretical Basis Document (ATBD): Water Quality},
  publisher = {National Ecological Observatory Network (NEON)},
  year      = {2021}
}

@misc{Neon_reaeration,
  doi       = {},
  url       = {https://data.neonscience.org/data-products/DP1.20190.001},
  author    = {{National Ecological Observatory Network (NEON)}},
  language  = {en},
  title     = {Reaeration field and lab collection (DP1.20190.001)},
  publisher = {National Ecological Observatory Network (NEON)},
  year      = {2021}
}

@inproceedings{ogasawara2010,
  title        = {Adaptive normalization: A novel data normalization approach for non-stationary time series},
  author       = {Ogasawara, Eduardo and Martinez, Leonardo C and De Oliveira, Daniel and Zimbr{\~a}o, Geraldo and Pappa, Gisele L and Mattoso, Marta},
  booktitle    = {The 2010 International Joint Conference on Neural Networks (IJCNN)},
  pages        = {1--8},
  year         = {2010},
  organization = {IEEE}
}

@manual{Rconduits,
  title  = {conduits: {CONDitional UI for Time Series normalisation}},
  author = {Puwasala Gamakumara and Priyanga Dilini Talagala and Rob J. Hyndman},
  year   = {2023},
  note   = {R package version 1.0.0},
  url    = {https://github.com/PuwasalaG/conduits}
}

@manual{Rsoftware,
  title        = {R: A Language and Environment for Statistical Computing},
  author       = {{R Core Team}},
  organization = {R Foundation for Statistical Computing},
  address      = {Vienna, Austria},
  year         = {2021},
  url          = {https://www.R-project.org/}
}

@article{seyam2014,
  title     = {The influence of accurate lag time estimation on the performance of stream flow data-driven based models},
  author    = {Seyam, Mohammed and Othman, Faridah},
  journal   = {Water resources management},
  volume    = {28},
  number    = {9},
  pages     = {2583--2597},
  year      = {2014},
  publisher = {Springer}
}

@manual{Stan,
  title  = {Stan Modeling Language Users Guide and Reference Manual},
  author = {{Stan Development Team}},
  year   = 2021,
  note   = {Version 2.28},
  url    = {https://mc-stan.org}
}

@article{vafaeipour2014,
  title     = {Application of sliding window technique for prediction of wind velocity time series},
  author    = {Vafaeipour, Majid and Rahbari, Omid and Rosen, Marc A and Fazelpour, Farivar and Ansarirad, Pooyandeh},
  journal   = {International Journal of Energy and Environmental Engineering},
  volume    = {5},
  number    = {2},
  pages     = {1--7},
  year      = {2014},
  publisher = {Springer}
}

@article{van2010,
  title     = {Nitrate response of a lowland catchment: On the relation between stream concentration and travel time distribution dynamics},
  author    = {Van der Velde, Y and De Rooij, GH and Rozemeijer, JC and Van Geer, FC and Broers, HP},
  journal   = {Water Resources Research},
  volume    = {46},
  number    = {11},
  year      = {2010},
  publisher = {Wiley Online Library}
}

@book{wanielista1997,
  title     = {Hydrology: Water quantity and quality control.},
  author    = {Wanielista, Martin and Kersten, Robert and Eaglin, Ron and others},
  year      = {1997},
  publisher = {John Wiley and Sons}
}

@article{wood2003TP,
  title     = {Thin plate regression splines},
  author    = {Wood, Simon N},
  journal   = {Journal of the Royal Statistical Society: Series B (Statistical Methodology)},
  volume    = {65},
  number    = {1},
  pages     = {95--114},
  year      = {2003},
  publisher = {Wiley Online Library}
}

@book{wood2017mgcvR,
  title     = {Generalized additive models: an introduction with R},
  author    = {Wood, Simon N},
  year      = {2017},
  publisher = {CRC press}
}

@article{xie2019,
  title     = {Normalization of large-scale behavioural data collected from zebrafish},
  author    = {Xie, Rui and Zhang, Mengrui and Venkatraman, Prahatha and Zhang, Xinlian and Zhang, Gaonan and Carmer, Robert and Kantola, Skylar A and Pang, Chi Pui and Ma, Ping and Zhang, Mingzhi and others},
  journal   = {Plos one},
  volume    = {14},
  number    = {2},
  pages     = {e0212234},
  year      = {2019},
  publisher = {Public Library of Science San Francisco, CA USA}
}

\end{document}